\documentclass[twocolumn,tighten]{aastex631}

\usepackage{pifont}
\newcommand{\cmark}{\ding{51}}%
\usepackage{bm}
\usepackage{soul}
\received{January 21, 2022}
\revised{February 21, 2022}
\accepted{February 28, 2022}
\watermark{Accepted}
\usepackage{multirow}
\usepackage{longtable,tabu}

\newcommand{\blu}[1]{{\textcolor[rgb]{0.0,0.0,0.0}{#1}}}

\newcommand{\be}{\begin{equation}}
\newcommand{\ee}{\end{equation}}
\newcommand{\bea}{\begin{eqnarray}}
\newcommand{\eea}{\end{eqnarray}}

\usepackage{amsmath}

\usepackage{xcolor}
\newcommand\crule[3][black]{\textcolor{#1}{\rule{#2}{#3}}}
\definecolor{color1}{HTML}{440154}
\definecolor{color2}{HTML}{481568}
\definecolor{color3}{HTML}{482677}

\definecolor{color18}{HTML}{B8DE29}
\definecolor{color19}{HTML}{DCE318}
\definecolor{color20}{HTML}{FDE725}
\shorttitle{Relativistic dense matter EOS and 
compatibility with NS observables: a Bayesian approach}
\shortauthors{Malik et al.}
\graphicspath{{./}{figures/}}

\begin{document}
\title{Relativistic description of dense matter equation of state and compatibility with neutron star observables: a Bayesian approach}

\correspondingauthor{Tuhin Malik}
\email{tuhin.malik@gmail.com, tuhin.malik@uc.pt}

\author[0000-0003-2633-5821]{Tuhin Malik}
\affiliation{CFisUC, Department of Physics, University of Coimbra, 3004-516 Coimbra, Portugal}

\author[0000-0002-5879-6262]{Márcio Ferreira}
\email{marcio.ferreira@uc.pt}
\affiliation{CFisUC, Department of Physics, University of Coimbra, 3004-516 Coimbra, Portugal}


\author[0000-0001-5032-9435]{B. K. Agrawal}
\email{bijay.agrawal@saha.ac.in}
\affiliation{Saha Institute of Nuclear Physics, 1/AF 
Bidhannagar, Kolkata 700064, India.}
\affiliation{Homi Bhabha National Institute, Anushakti Nagar, Mumbai 400094, India.}

\author[0000-0001-6464-8023]{Constan\c ca Provid\^encia}
\email{cp@uc.pt}
\affiliation{CFisUC, Department of Physics, University of Coimbra, 3004-516 Coimbra, Portugal}

\begin{abstract}
The general behavior of the nuclear equation of state (EOS), relevant for the description of neutron stars (NS), is studied within a  Bayesian approach applied to a set of models based on a density dependent relativistic mean field description of nuclear matter. The EOS is subjected to a  minimal number of constraints based on nuclear saturation properties and  the low density pure neutron matter EOS  obtained from a precise next-to-next-to-next-to-leading order (N$^{3}$LO) calculation in chiral effective field theory ($\chi$EFT). The posterior distributions of the model parameters obtained under these minimal constraints are employed to construct the distributions of various nuclear matter properties and NS properties such as radii, tidal deformabilites, central energy densities and speeds of sound etc. We found that 90\% confidence interval (CI) for allowed NS mass - radius relationship and tidal deformabilites are compatible with GW170817 and recent NICER observations, without invoking the exotic degrees of freedom.
\blu{A central speed-of-sound of the order of $\sqrt{2/3}$ $c$ is obtained. The maximum neutron star mass allowed by the model is  2.5$M_\odot$.
}

\end{abstract}

\section{Introduction} \label{sec:intro}
{Neutron stars (NS), observed as pulsars are one of the densest and  most compact objects in the universe. The core of such compact objects is believed to contain matter at few times nuclear saturation density ($\rho_0 =2.7 \times 10^{14}$ g/cm$^3$) \cite{book.Glendenning1996,book.Haensel2007,Rezzolla:2018jee}}. 
It is the ideal cosmic laboratory to test our present knowledge of the mysterious behavior of matter under extreme densities. The existence of NS was first hypothesized by Lev Landau, see \cite{Yakovlev:2012rd}, and by Walter Baade and Fritz Zwicky  in 1933 \cite{Baade1934a,Baade1934b}. However, Jocelyn Bell and her Ph.D. advisor A. Hewish first observed neutron stars in 1967 with the discovery of radio pulsars \cite{Bell1968}. A detail history on the origin of NS can be found in Ref \cite{Brecher1999}. The NS properties namely, the maximum mass, radii, moments of inertia, and tidal Love numbers of neutron stars, all of which are accessible to observation can be a significant probe to reduce the uncertainty on theoretical models of NS over the decades. The high mass pulsars like PSR ~J1614-2230 ($M = 1.908 \pm~ 0.016 M_{\odot}$)  \cite{Demorest2010,Fonseca2016,Arzoumanian2017}, PSR~ J0348 - 0432 ($M = 2.01 \pm~ 0.04~ M_{\odot}$) \cite{Antoniadis2013},  PSR J0740+6620 ($M = 2.08 \pm~ 0.07~ M_{\odot}$  \cite{Fonseca:2021wxt} and very recently J1810+1714 with a mass $M = 2.13 \pm~ 0.04~ M_{\odot}$  \cite{Romani:2021xmb} have drawn attention to the theory of nuclear interactions at high density. The high-precision X-ray space missions, such as the NICER (Neutron star Interior Composition ExploreR) have already shed some light in this direction. Of late, NICER has come up with one measurement of the radius $12.71_{-1.19}^{+1.14}$ km and mass $1.34_{-0.16}^{+0.15}$ M$_\odot$  for the pulsar PSR J0030+0451 \cite{Riley:2019yda}, and other independent analysis shows that the radius is $13.02_{-1.06}^{+1.24}$ km and the mass $1.44_{-0.14}^{+0.15}$ M$_\odot$ \cite{Miller:2019cac}. 
The recent measurement of the equatorial circumferential radius of the  pulsar PSR J0740+6620 with  mass $M=2.072_{-0.066}^{+0.067}$ M$_\odot$ and   $R=12.39^{+1.30}_{-0.98}$ km (68 $\%$ CI) \cite{Riley:2021pdl},
by NICER group will play a important role in this domain. The empirical estimates of  the radius of a canonical NS ($M = 1.4 M_{\odot}$) is  $R_{1.4} = (11.9 \pm 1.22)~ $ km  according to \cite{Lattimer2013}. Recently, from the simultaneous analysis of  NICER and  XMM-Newton X-ray observations an estimation of $12.45 \pm 0.65$~km at 68\% CI was obtained for a $1.4 M_{\odot}$ star.

{The internal structure of the NS depends on the hydrostatic equilibrium between the inward gravitational pull of matter and the outward neutron degeneracy pressure. General Relativity allows us to calculate the internal structure of NS. The first NS model was calculated by Oppenheimer \& Volkoff \cite{TOV2} using the exact form of the equations of hydrostatic equilibrium in General Relativity, which they derived simultaneously with Tolman \cite{TOV1} from the Einstein equations. {To solve NS structure equations, i.e., Tolman-Oppenheimer-Volkoff (TOV) equations, one needs the theory of the behavior of matter under extreme conditions, i.e., the theory of the infinite nuclear matter equation of state (EOS)}. The knowledge of the nuclear many body theory is necessary for obtaining the nuclear matter EOS. 
In general, phenomenological models for nuclear EOS can be broadly categorized into two groups: (i) the relativistic and (ii) the non-relativistic models. Although, non-relativistic methods have been extremely successful in the description of nucleons inside atomic nuclei (finite nuclei), for infinite dense nuclear matter one needs to consider relativistic effects and assure that the speed of sound is always below the speed of light. A different approach treats the  nuclear interaction in a relativistic framework \cite{Serot1984}. Relativistic mean field  (RMF) models are specially adequate to describe high density matter as the one occurring inside NS, besides also describing finite nuclei.
In fact, RMF models  successfully  deal with the inclusion of many body effects in the description of finite nuclei and infinite nuclear matter via the exchange of mesons ($\sigma$, $\omega$ and $\varrho$). In order to  describe nuclear properties two different approaches have been developed: non-linear meson terms are included in the Lagrangian density in order to describe adequately the density dependence of the EOS and symmetry energy \cite{Boguta1977,Mueller1996,Steiner2004,Todd-Rutel2005}; the non-linearities are described introducing density dependent coupling parameters and avoiding the introduction of non-linear mesonic terms \cite{Typel1999,Typel2009,Lalazissis2005}. These models are phenomenological and need to be constrained by experimental or observational data. However, the presently existing data from the laboratory are obtained from nuclei that have {a proton fraction not much smaller than 0.4} and the densities attained are normally of the order of the saturation density or below. This imposes big limitations in these models: their extension to high densities and/or isospin asymmetries has to be taken with care.

Recently, several EOS metamodels constrained by {\it ab-initio} theoretical calculations for both low and high density  have been proposed: nucleon-nucleon chiral potentials for the low density neutron and nuclear matter \cite{Hebeler2013,Drischler2015}  and  perturbative Quantum Chromodynamics for asymptotically high-density regimes \cite{Kurkela2009}. In order to account for all possible EOS compatible with these two constraints, the EOS at the two extreme densities are connected  using a piecewise polytropic interpolation, a speed-of-sound interpolation or a spectral interpolation, and  causality is imposed when necessary \cite{Lindblom2012,Kurkela:2014vha,Most:2018hfd,LopeOter:2019pcq,Annala2019,Annala:2021gom}. {Of late, a nonparametric inference of the NS EOS has also been proposed based on Gaussian processes (GPs) \cite{Essick2019} or using machine learning techniques \cite{Han:2021kjx}.} {However, such EOS models have strong limitations because they  do not assume any kind of composition of matter in the intermediate density regime.} Other approach has been considered that also span an acceptable NS mass-radius domain such as a Taylor expansion parametrization of the EOS \cite{Margueron2018a,Margueron2018b,Zhang2018,Ferreira2021,Ferreira:2021pni}. {The recovery of the nuclear matter properties from the $\beta$-equilibrium EOS has proven to be impossible without the knowledge of the compositions {or symmetry energy at high densities} \cite{Tovar2021,Imam2021,Mondal2021} {or the knowledge of the EOS of symmetric nuclear matter along with compositions  \cite{Essick2021}.}}

{The aim of the present study is to generate a set of models 
{using microscopic approach based on relativistic description of
hadrons through their density-dependent coupling with mesons (DDH),}
constrained by existing observational, theoretical and experimental data. This has as a basic hypothesis that NS matter is nucleonic matter with electrons and muons. The approach has the great advantage of being a causal description of matter and it will not be necessary to impose a speed of sound below the speed of light.  The framework will, however, be easily extended to allow the inclusion of other degrees of freedom such as hyperons or a deconfinement transition. Considering only the simplest composition will allow us to evaluate how much the existing constraints require the introduction of exotic degrees of freedom, to explain NS. {In \cite{Thi:2021jhz}, the authors have concluded within a meta-model description that the present NS observations are compatible with what they call the ``nucleonic hypotheses'', i.e. nucleonic and leptonic degrees of freedom are sufficient to explain the data. }

The advantages of a DDH approach with respect to an approach with constant couplings is that it accounts for quantal fluctuations  of the baryon fields even in the ground state \cite{Lenske:1995wyj}. The rearrangement self-energies to the baryon field equations
are responsible by such effects. Over the decades, several formulations of density-dependent couplings have been studied \cite{Fritz:1994ww,Marcos:1989zz,Haddad:1993zz} employing the usual field equations and definitions of self-energies. However, a closer inspection suggests that not all these models were consistent \cite{Lenske:1995wyj,Fuchs:1995as}. To obtain a Lorentz-invariant Lagrangian and covariant field equations from the Euler-Lagrange equations, the density dependence of the couplings has to be a Lorentz-scalar functional of the baryon fields. The development of a DDH parameterization that  simultaneously describes the properties of nuclear matter and finite nuclei has been very successful \cite{Typel1999,Typel2009,Lalazissis2005}. This model  also allows a reasonable extrapolation to extreme conditions of isospin and density.}

{In the present work, we perform a detailed statistical analysis of
the parameters of a DDH description of nuclear matter  within an Bayesian approach considering a given set of fit data related with the nuclear saturation properties,
the pure neutron matter EOS calculated from a precise N$^3$LO calculation
in $\chi$EFT and  the lower bound of observed two solar mass NS. We introduce a  density dependence of the couplings of iso-vector and iso-scalar fields so that for each field only two parameters are necessary: the dependence of the couplings on the density is compatible with Dirac-Brueckner-Hartree-Fock calculations and similar to the one introduced in
 \cite{Typel1999}. Two sets of models will be generated that only differ on the constraints used to fit the model parameters. Once the sets are built, a  detail statistical analysis of nuclear matter parameters (NMPs), also some that are not directly accessible in laboratory experiments, and of neutron star properties will be performed. A comparison of the main results with the ones obtained within other approaches, such as the a meta-model description of NS matter, will be discussed. It will be shown that the set of models constrained by $\chi$EFT pure neutron matter calculations \cite{Hebeler2013} and some saturation nuclear matter properties are totally compatible with NICER and GW170817 observations. Besides, it will also be shown that the higher order NMP compatible with NS constraints may differ from results determined from  Taylor expansion EOS.

{The paper is organized as follows, In Section \ref{model}, the field theoretical  DDH model for the  EOS at zero and finite temperatures is briefly reviewed, followed by a brief description of Bayesian estimation of model parameters  in Section \ref{bayes}. The results of our calculation are discussed in Section \ref{results}. Section \ref{con} contains the summary and conclusions.}

\section{FORMALISM}
In this section, the RMF framework used to generate the set of models that will be applied in the present study is introduced. A brief review of the Bayesian  approach undertaken to estimate the model parameters will be presented.

\subsection{Model\label{model}}
The calculation of the nuclear EOS boils down to a problem of the theoretical modeling of the nuclear interactions. In a phenomenological approach, the effective interactions among nucleons can be modeled within a relativistic mean field framework with an effective Lagrangian involving baryon and meson fields. The force between two nucleons is realized by the exchange of mesons in this framework. The $\sigma$ meson creates a strong attractive central force and influences the spin-orbit potential, on the other hand, the $\omega$-meson is responsible for the repulsive short range force. The isovector $\varrho$ meson is included to distinguish between neutrons and protons, and introduce the isospin symmetry and  independence of the nuclear force. The Lagrangian including the nucleon field, the $\sigma$, $\omega$ and $\varrho$ mesons and their interactions can
be written as,
\begin{equation}
\begin{aligned}
\mathcal{L}=& \bar{\Psi}\Big[\gamma^{\mu}\left(i \partial_{\mu}-\Gamma_{\omega} A_{\mu}^{(\omega)}-
\Gamma_{\varrho} {\boldsymbol{\tau}} \cdot \boldsymbol{A}_{\mu}^{(\varrho)}\right) \\
&-\left(m-\Gamma_{\sigma} \phi\right)\Big] \Psi 
+ \frac{1}{2}\Big\{\partial_{\mu} \phi \partial^{\mu} \phi-m_{\sigma}^{2} \phi^{2} \Big\} \\
&-\frac{1}{4} F_{\mu \nu}^{(\omega)} F^{(\omega) \mu \nu} 
+\frac{1}{2}m_{\omega}^{2} A_{\mu}^{(\omega)} A^{(\omega) \mu} \\
&-\frac{1}{4} \boldsymbol{F}_{\mu \nu}^{(\varrho)} \cdot \boldsymbol{F}^{(\varrho) \mu \nu} 
+ \frac{1}{2} m_{\varrho}^{2} \boldsymbol{A}_{\mu}^{(\varrho)} \cdot \boldsymbol{A}^{(\varrho) \mu},
\end{aligned}
\label{lagrangian}
\end{equation}
where  $\Psi$ is the Dirac spinor for spin $\frac{1}{2}$ particles, and, in the present calculation, describes a nucleon doublet (neutron and proton) with bare mass $m$.
$\gamma^\mu $ and $\boldsymbol{\tau}$ are the Dirac matrices and the Pauli matrices, respectively. The vector meson field strength tensors are given by   $F^{(\omega, \varrho)\mu \nu} = \partial^ \mu A^{(\omega, \varrho)\nu} -\partial^ \nu A^{(\omega, \varrho) \mu}$. The $\Gamma_{\sigma}$, $\Gamma_{\omega}$ and $\Gamma_{\varrho}$ are the coupling constants of nucleons to the meson fields $\sigma$, $\omega$ and $\varrho$, respectively, and the corresponding meson masses are $m_\sigma$, $m_\omega$ and $m_\varrho$. A DDH model is considered with
nucleon-meson   density-dependent coupling parameters in the form of}
\begin{equation}
  \Gamma_{M}(\rho) =\Gamma_{M,0} ~ h_M(x)~,\quad x = \rho/\rho_0~,
\end{equation}
where the  density $\rho$ is the baryonic density, the $\Gamma_{M,0}$ is the couplings at saturation density $\rho_0$ and $M \in \{ \sigma, \omega, \varrho \}$. For the isoscalar couplings,  in the present study  the function $h_M$ is given by
\begin{equation}
h_M(x) = \exp[-(x^{a_M}-1)]
\label{hm1}
\end{equation}
and  the isovector coupling has the form proposed in \cite{Typel1999}
\begin{equation}
h_\varrho(x) = \exp[-a_\varrho (x-1)] ~.
\label{hm2}
\end{equation}
{The parametrization defined in Eq. (\ref{hm1}) introduces only one extra parameter for each coupling, similarly to the $\varrho$-meson coupling,   and was chosen so that the $\sigma$ and $\omega$-nucleon couplings may have a dependence on the density as predicted from Dirac-Br\"uckner-Hartree-Fock calculations \cite{TerHaar1987,Brockmann1990,Typel1999}, for $\rho\gtrsim 0.04$ fm$^{-3}$. This range of densities is adequate to describe the NS core EOS.}

In the following, we use the  mean field approximation, and we consider that the system is formed of static uniform matter in its ground state. The mesonic fields are replaced by their expectation value $\langle \sigma \rangle$ and $\langle  A^{(\omega, \varrho)}_\mu\rangle$, and quantum fluctuation are neglected. In static uniform matter, the source {densities and} currents  $\bar{\psi}(x) \psi(x)$ and $\bar{\psi}(x) \gamma^{\mu} \psi(x)$ are independent of $x$. Besides, only the time-like components of vector fields $\omega_0$ and the third isospin component of the $\varrho$ field $\varrho_3^0$  survive. The Euler-Lagrange equations of all the fields are in the mean field approximation 
\begin{eqnarray}
m_{\sigma}^{2} \sigma= \Gamma_{\sigma} \bar{\psi} \psi, \\
\quad m_{\omega}^{2} \omega_{0}= \Gamma_{\omega} \bar{\psi} \gamma_{0} \psi, \\
\quad m_{\varrho}^{2} \varrho_{3}^0=\frac{1}{2} \Gamma_{\varrho} \bar{\psi} \gamma_{0} \tau_{3} \psi
\end{eqnarray}
The nucleon number density $\rho=<\bar{\psi} \gamma_{0} \psi>$ and scalar density $\rho_s=<\bar{\psi} \psi>$
at zero temperature are defined as,
\begin{eqnarray}
\rho=\frac{\gamma}{2\pi^{2}} \sum_{B=p,n}\int_{0}^{k_{F_B}} 
k^2\,d k, \\
\rho_{s}=\frac{\gamma}{2 \pi^{2}} \sum_{B=p,n}\int_{0}^{k_{F_B}} 
\frac{m^{*} k^2}{\sqrt{m^{* 2}+k^{2}}} \,dk,
\end{eqnarray}
where $k_{F_B}$  is  the  Fermi momentum of nucleon $B$ and $\gamma$ is the spin degeneracy factor. 
The effective nucleon mass is $m^{*}=m - \Gamma_{\sigma} \sigma$ and the nucleon $B$ chemical potential is given by $\mu_B=\nu_{B}+ \Gamma_{\omega} \omega_{0}+ \Gamma_{\varrho} \tau_{3 B} \varrho_{3}^0 +\Sigma^{r} $, where $\tau_{3 B}$ is the isospin projection and  the rearrangement term $\Sigma^{r}$ takes care of many-body effects in nuclear interaction \cite{Typel1999}, and assures thermodynamic consistency. It arises due to the density-dependence of the couplings and is expressed as
\begin{equation}
\Sigma^{r}=\sum_{B=n,p}\left[-\frac{\partial \Gamma_{\sigma}}{\partial \rho_{B}} \sigma \rho_{sB}+\frac{\partial \Gamma_{\omega}}{\partial \rho_{B}} \omega_{0} \rho_{B}+\frac{\partial \Gamma_{\varrho}}{\partial \rho_{B}} \tau_{3 B} \rho_{3}^0 \rho_{B}\right].
\end{equation}
The energy density is defined as,
\begin{eqnarray}
\varepsilon&=&\frac{1}{\pi^{2}} \sum_{B={n},{p}} \int_{0}^{k_{F_B}} k^{2} \sqrt{k^{2}+m^{* 2}} d k + \frac{1}{2} m_{\sigma}^{2} \sigma^{2} \nonumber \\
&+&\frac{1}{2} m_{\omega}^{2} \omega_{0}^{2}+\frac{1}{2} m_{\varrho}^{2} (\varrho_{3}^0)^{2}+ {\varepsilon_{lep}},
\end{eqnarray}
{where the last term describes the leptonic (electrons and muons) contribution.}
The pressure P can be derived from the energy density using the Euler relation,
\begin{equation}
P={\sum_{i=n,p,e,\mu}} \mu_{i} \rho_{i}-\varepsilon,
\end{equation}
{where $\mu_i$ and $\rho_i$ are, respectively, the chemical potential and the number density of particle $i$.}

In the core, the star is mainly composed of neutrons with  very 
high momentum states. $\beta$-decay establishes an equilibrium between neutrons, protons, electrons and muons
%
\bea 
n \leftrightarrow p + e^- + \bar{\nu}, \\
n + \nu \leftrightarrow p + e^- ,\\
{\mu\leftrightarrow e^-+ \nu_\mu+\bar\nu_e,}
\eea 
and muons ($\mu$) will appear when the 
chemical potential of the electrons reaches the muon rest mass ($m_{\mu} = 106$ MeV). {In a cold catalyzed NS, the wavelength of neutrinos is much larger than the star radius and they escape.}
Therefore, the $\beta$-equilibrium condition is given as,
\bea 
\label{ch02}
\mu_n=\mu_p+\mu_e \quad {\rm and}\qquad\mu_e=\mu_\mu.
\eea 
For a given baryon density ($\rho=\rho_n+\rho_p$), the charge neutrality imposes,
\bea 
\label{ch01}
\rho_p=\rho_e+\rho_{\mu}.
\eea 

{In order to obtain the NS properties, it is necessary to match the crust EOS to the core EOS. For the outer crust the Bethe-Pethick-Sutherland (BPS) EOS is chosen. 
The outer crust and the core are joined using the polytropic 
form \cite{Carriere:2002bx} $p(\varepsilon)=a_1 + a_2 \varepsilon^{\gamma}$, where the parameters $a_1$ and $a_2$ are determined in such a way that 
the EOS for the inner crust  matches with the outer crust at one end ($\rho=10^{-4}$ fm$^{-3}$) and with the core at the other end ($\rho=0.04$ fm$^{-3}$). The polytropic index $\gamma$ is taken to be equal to $4/3$.
} 
\blu{This approximation will introduce an uncertainty on the radius of the low mass NS as shown in \cite{Fortin2016,Pais:2016xiu}, see also the recent studies \cite{Lopes:2020xlf,Rather:2020gja}. In \cite{Fortin2016} several matching procedures have been tested and it was shown that the uncertainty could be as high as 1 km. One of the methods that introduced a small uncertainty considered a matching to the outer core at a density 0.01~fm$^{-3}$. The justification being the fact the inner crust EOS does not differ much from the homogeneous EOS for densities close to the transition to the core, as clearly seen in Fig. 5 of Ref. \cite{Avancini2008} for DDH models. We, therefore, believe that our approximation will introduce an uncertainty  in the radius of a 1.4~$M_\odot$ star  that is at most of the order of $100-200$m for models with a symmetry energy compatible with the $\chi$EFT PNM EOS, see Table 1 of \cite{Fortin2016}, and smaller for larger masses. We have estimated for the five models given in the supplementary material, DDBl, DDBm, DDBu1, DDBu2 and DDBx, the uncertainty on the radius of a 1.4$M_\odot$ star introduced with our approach. For these five models  we have calculated the inner crust within an approach that includes the surface energy and Coulomb field after minimization \cite{Avancini:2008zz} and we have obtained a difference of $\lesssim  10$~m (DDBl and DDBm), $\lesssim 100$~m (DDBu2), $\approx 150$~m (DDBu1)  and  $\lesssim 200$~m for one of the extreme EOS with $K_0=300$~MeV, DDBx.}

To a good approximation, the EOS of nuclear matter can be decomposed into two parts, (i) the EOS for symmetric nuclear matter $\epsilon(\rho,0)$ (ii) a term involving the symmetry energy coefficient $S(\rho)$ and the asymmetry $\delta$,
\bea
 \epsilon(\rho,\delta)\simeq \epsilon(\rho,0)+S(\rho)\delta^2,
 \label{eq:eden}
\eea 
{where  $\epsilon$ is the energy per nucleon at a given density  $\rho$  and isospin asymmetry} $\delta=(\rho_n-\rho_p)/\rho$.
We can recast the EOS in terms of various bulk nuclear matter properties {of order $n$ at saturation density: (i) for the symmetric nuclear matter, the energy per nucleon $\epsilon_0=\epsilon(\rho_0,0)$ ($n=0$), the incompressibility coefficient $K_0$ ($n=2$), the
skewness  $Q_0$ ($n=3$),  and  the kurtosis $Z_0$ ($n=4$), respectively, given by
\begin{equation}
X_0^{(n)}=3^n \rho_0^n \left (\frac{\partial^n \epsilon(\rho, 0)}{\partial \rho^n}\right)_{\rho_0}, \, n=2,3,4;
\label{x0}
\end{equation}
(ii) for the symmetry energy,  the symmetry energy at saturation 
 $J_{\rm sym,0}$ ($n=0$), 
\begin{equation}
J_{\rm sym,0}= S(\rho_0)=\frac{1}{2} \left (\frac{\partial^2 \epsilon(\rho,\delta)}{\partial\delta^2}\right)_{\delta=0},
\end{equation}
the slope $L_{\rm sym,0}$ ($n=1$),  the curvature $K_{\rm sym,0}$ ($n=2$),  the skewness $Q_{\rm sym,0}$ ($n=3$), and  the kurtosis $Z_{\rm sym,0}$ ($n=4$), 
respectively, defined as
\begin{equation}
X_{\rm sym,0}^{(n)}=3^n \rho_0^n \left (\frac{\partial^n S(\rho)}{\partial \rho^n}\right )_{\rho_0},\, n=1,2,3,4.
\label{xsym}
\end{equation}
}

\subsection{Bayesian estimation of Model Parameters\label{bayes}}
{A Bayesian parameter estimation approach, enables one to carry out a detailed statistical analysis of the parameters of a model for a given set of fit data \cite{Wesolowski:2015fqa,Furnstahl:2015rha,Ashton2019,Landry:2020vaw}. In this technique, the basic rules of probabilistic inference are used to update the probability for a hypothesis under the available evidence or information according to  Bayes' theorem. The posterior distributions of the model parameters $\theta$ in Bayes’ theorem can be written as
\begin{equation}
P(\bm{\theta} |D ) =\frac{{\mathcal L } (D|\bm{\theta}) P(\bm {\theta })}{\mathcal Z},\label{eq:bt}
\end{equation}
where $\bm{\theta}$  and $D$ denote the set of model parameters and the fit data.  $P(\bm {\theta })$ in 
Eq. (\ref{eq:bt}) is the prior for the  model parameters and $\mathcal Z$ is the evidence. The type of prior can be chosen with the preliminary knowledge of the model parameters. One can choose it to be a uniform prior, which has been used as a baseline for many analyses. The $P(\bm{\theta} |D )$ is the joint posterior distribution of the parameters, $\mathcal L (D|\bm{\theta})$ is the likelihood function.
The posterior distribution  of a given parameter can be obtained by marginalizing $P(\bm{\theta} |D )$ over  the remaining parameters. The marginalized posterior distribution for a  parameter $\theta_i$
is obtained as,
\begin{equation}
 P (\theta_i |D) = \int P(\bm {\theta} |D) \prod_{k\not= i }d\theta_k. \label{eq:mpd}
\end{equation}
We use a Gaussian likelihood function defined as, 
\bea
{\mathcal L} (D|\bm{\theta})&=&\prod_{j} 
\frac{1}{\sqrt{2\pi\sigma_{j}^2}}e^{-\frac{1}{2}\left(\frac{d_{j}-m_{j}(\bm{\theta)}}{\sigma_{j}}\right)^2}. 
\label{eq:likelihood}  
\eea
Here the index $ j$ runs over all the data, $d_j$ and $m_j$ are the data and corresponding model values, respectively.  The $\sigma_j$ are the adopted uncertainties.} 
{The Markov Chain Monte Carlo (MCMC) is commonly employed  for Bayesian parameter estimation. This algorithm jumps to a new set of parameters from starting parameters with a probability proportional to the ratio of the two points. It is a powerful algorithm for high dimensionality problem. However, the MCMC has its own problems with convergence. To overcome the problem of MCMC, a different Monte Carlo algorithm, Nested Sampling, was first proposed in Ref. \cite{Skilling2004}. In Nested Sampling,  the posterior is broken into many nested “slices” with starting {\it"n-live"} points,  samples are generated from each of them and then recombined  to reconstruct the original distribution.  In a Dynamic Nested Sampling the procedure is similar but the {\it"n-live"} varies dynamically. We have implemented both the Nested Sampling and the Dynamic Nested Sampling algorithm in the Bayesian Inference Library (BILBY) \cite{Ashton2019} to populate the posterior distribution of Eq. (\ref{eq:bt}) by invoking a {\it Pymultinest} sampler \cite{Buchner:2014nha,buchner2021nested} and a {\it Dynesty} sampler  \cite{Speagle2019}, respectively.}

{We generate samples for starting 3000 {\it"n-live"} points with both  samplers, separately. The {\it Pymultinest} selects around 14000 final models by calling $5.78 \times 10^5$ models and the {\it Dynesty} selects around 13000 final models by calling $5 \times 10^7$ models. The evidence obtained in both  samplers are similar. In the next section, we will present the results sets obtained in {\it Pymultinest}. }

\bigskip

\section{Results \label{results}}
{In this section, we study the dense matter EOS relevant for NS in the DDH framework as {briefly outlined} in Sec. \ref{model}. A detailed statistical analysis of the DDH model parameters, namely $\Gamma_{\sigma,0}$, $\Gamma_{\omega,0}$, $\Gamma_{\varrho,0}$, $a_\sigma$, $a_\omega$ and $a_\varrho$, is done within a Bayesian parameter estimation approach considering a given set of fit data related with the nuclear saturation properties, the pure neutron matter EOS calculated from a precise N$^3$LO calculation in $\chi$EFT and the lowest bound of NS observational maximum mass. With the  marginalized posterior distributions obtained for the DDH parameters, we  perform a statistical analysis of nuclear matter parameters and the NS properties. The marginalized posterior distributions of the DDH parameters, applying a  Bayesian estimation of  the model parameters, requires the definition of  the likelihood, of the fit data and of the priors for the model parameters. The likelihood has been defined in Sec. \ref{bayes}, see Eq. (\ref{eq:likelihood})}.
\begin{table*}[]
 \caption{The list of data/constraints considered in the Bayesian {estimation} of the model parameters \blu{which generate the DDB set}. The $\epsilon_0$ is the binding energy per nucleon, $K_0$ {the} incompressibility coefficient and $J_{\rm sym,0}$ the symmetry energy  evaluated at the nuclear saturation density  $\rho_0$. The nuclear saturation properties are listed including an 1$\sigma$ uncertainty. The PNM indicates the pressure of pure neutron matter \blu{for the densities 0.08, 0.12 and 0.16~fm$^{-3}$} from N$^3$LO calculation in $\chi$EFT \cite{Hebeler2013}. We consider {2 $\times$} N$^3$LO data in the likelihood of the present calculation.}
  \label{tab1}
 \setlength{\tabcolsep}{5.5pt}
      \renewcommand{\arraystretch}{1.1}
\begin{tabular}{ccccc}
\toprule
\multicolumn{5}{c}{Constraints}                                                        \\
\multicolumn{2}{c}{Quantity}                     & Value/Band  & Ref   & DDB   \\ \hline
\multirow{3}{*}{\shortstack{NMP \\  {[}MeV{]} }} 
& $\rho_0$ & $0.153\pm0.005$ & \cite{Typel1999}  & \cmark  \\
& $\epsilon_0$ & $-16.1\pm0.2$ & \cite{Dutra:2014qga} & \cmark  \\
                               & $K_0$           & $230\pm40$   & \cite{Shlomo2006,Todd-Rutel2005}  & \cmark  \\
                              & $J_{\rm sym, 0}$           & $32.5\pm1.8$  & \cite{Essick:2021ezp} & \cmark  \\
                              
                               &                 &                & &                                                   \\
  \shortstack{PNM \\ {[}MeV fm$^{-3}${]}}                  & $P(\rho)$       & $2\times$ N$^{3}$LO    & \cite{Hebeler2013} & \cmark  \\
\shortstack{NS mass \\ {[}$M_\odot${]}}        & $M_{\rm max}$   & $>2.0$     &  \cite{Fonseca:2021wxt}     & \cmark \\ 
\toprule
\end{tabular}
\end{table*}
{We consider a minimal set of fit data, referred hereafter as DDB set, see Table \ref{tab1}. The data sets contain \blu{four} empirical nuclear saturation properties, the low density pressure for pure neutron matter \blu{at three different densities, in particular, 0.08, 0.12 and 0.16 fm$^{-3}$}, obtained from $\chi$EFT \cite{Hebeler2013} 
and the lowest bound of the neutron star maximum mass observational constraint.
The \blu{four} empirical nuclear saturation properties are: $\rho_0$ the nuclear saturation density, $\epsilon_0$  the binding energy per nucleon, $K_0$ {the} incompressibility coefficient and $J_{\rm sym,0}$ {the} symmetry energy coefficient all defined at the nuclear saturation density $\rho_0$. 
\blu{The range of values considered  for $J_{\rm sym,0}$  is the one defined by      the $\chi$EFT marginalized  values obtained in  \cite{Essick:2021ezp} considering four independent calculations \cite{Hebeler:2009iv,Tews:2012fj,Lynn:2015jua,Drischler:2017wtt} with equal weights.
{The N$^{3}$LO bound for PNM pressure restricts the symmetry energy within a very narrow range and to have a broader range for the symmetry energy, we consider a $2 \times$ N$^{3}$LO  uncertainty band for DDB set.}}
In Table \ref{tab2}, we show the prior set P of the DDH model parameters. The uniform prior has been taken with a reasonable boundary. We initially do a random sampling test with the Latin hypercube sampling (LHS) \cite{LHS1} to get the overall idea about a reasonable boundary of the parameter space, i.e. the sub-domain for which we get a physical equation of state. {It should be referred that the Nuclear Matter Parameters (NMPs) that result from the DDH parameters prior span also  a reasonable wide range} {of the domain of acceptable values for these parameters.}
\begin{table}[]
\caption{The considered uniform prior distributions (P) of the DDH model parameters. The parameters ’min’ and ’max’ denote the minimum and maximum values for the uniform distribution.}
\label{tab2}
\setlength{\tabcolsep}{15.5pt}
\renewcommand{\arraystretch}{1.1}
\begin{tabular}{cccc}
\toprule
\multirow{2}{*}{No} & \multirow{2}{*}{Parameters}  & \multicolumn{2}{c}{P} \\ \cline{3-4} 
                    &                                & min  & max  \\ \hline
1                   & $\Gamma_{\sigma,0}$                              & 7.5  & 13.5   \\
2                   & $\Gamma_{\omega,0}$                              & 8.5  & 14.5   \\
3                   & $\Gamma_{\varrho,0}$                        & 2.5    & 8.0  \\
4                   & $a_{\sigma}$                               & 0.0    & 0.30 \\
5                   & $a_{\omega}$                             & 0.0    & 0.30 \\
6                   & $a_{\varrho}$                            & 0.0 & 1.30    \\ \hline
\end{tabular}
\end{table}

\begin{figure}
\includegraphics[width=0.45\textwidth]{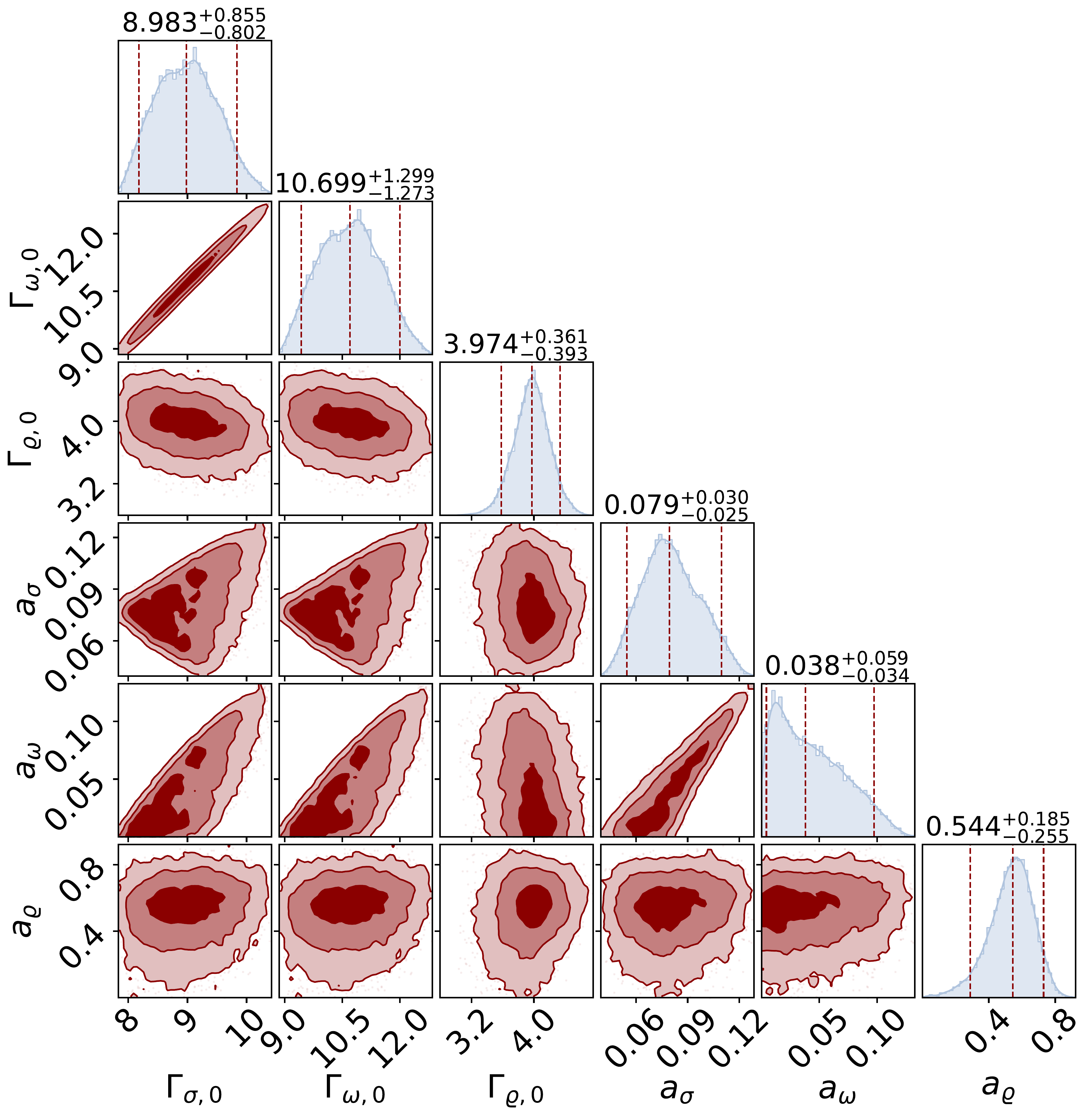}
\caption{Corner plots for the marginalized posterior distributions of our DDH model parameters. The results are obtained for the DDB  set with the prior set P of Table \ref{tab2}. One dimensional posterior distributions are given along the diagonal plots. The vertical lines indicate the 90\% min, median and 90\% max confidence interval of the model parameters, respectively. The confidence ellipses  for two dimensional posterior distributions are plotted with 1$\sigma$, 2$\sigma$ and 3$\sigma$ confidence intervals.
\label{T:fig1}} 
\end{figure}
{In Fig. \ref{T:fig1}, we show the corner plots for the  marginalized posterior distributions of the DDH model parameters $\Gamma_{\sigma,0}$, $\Gamma_{\omega,0}$, $\Gamma_{\varrho,0}$, $a_{\sigma}$, $a_{\omega}$ and $a_{\varrho}$, corresponding to the uniform prior set P presented in Table \ref{tab2} for data set DDB. The number of final sample parameters corresponding to the posterior sets are around {fourteen thousand}. The plots along the diagonal on the figure compare the one dimensional marginalized posterior distribution of individual parameters obtained for DDB  set. The vertical lines indicate the 90\% min, median and 90\% max credible interval (CI) of the distributions, respectively. The CI for the 2D marginalized posterior distributions are plotted with $1\sigma$, $2\sigma$ and $3\sigma$ CIs. The elliptical nature of the 2D CI for a few number of parameters indicate the correlations existing among those parameters, while a  circular nature indicates  no correlations. For example, as can be seen from the figure,  the parameters $\Gamma_{\sigma,0}$ and $\Gamma_{\omega,0}$ {as well as the parameters  $a_{\sigma}$ and $a_{\omega}$} are highly correlated  due to the nuclear binding energy at saturation imposed in DDB  set. It is to be noted, that $a_{\sigma}$ and $a_{\omega}$ determine the degree of non linearity in the iso-scalar part and $a_{\varrho}$ in the iso-vector part of the EOS at high density. In Table \ref{tab3} we list the median value and 68\% (90\%) CI for all model parameters obtained for DDB  set. The DDB set results in hard enough EOS by having in average a small $\sigma$-coupling, responsible for the description of attractive component of the nuclear force, a large $\rho$-meson coupling, responsible for the symmetry energy, and small parameters $a_i$ which avoid that the couplings of the vector mesons, that predominate at high densities, reduce too fast with density.}
\begin{figure}[t]
\includegraphics[width=0.45\textwidth]{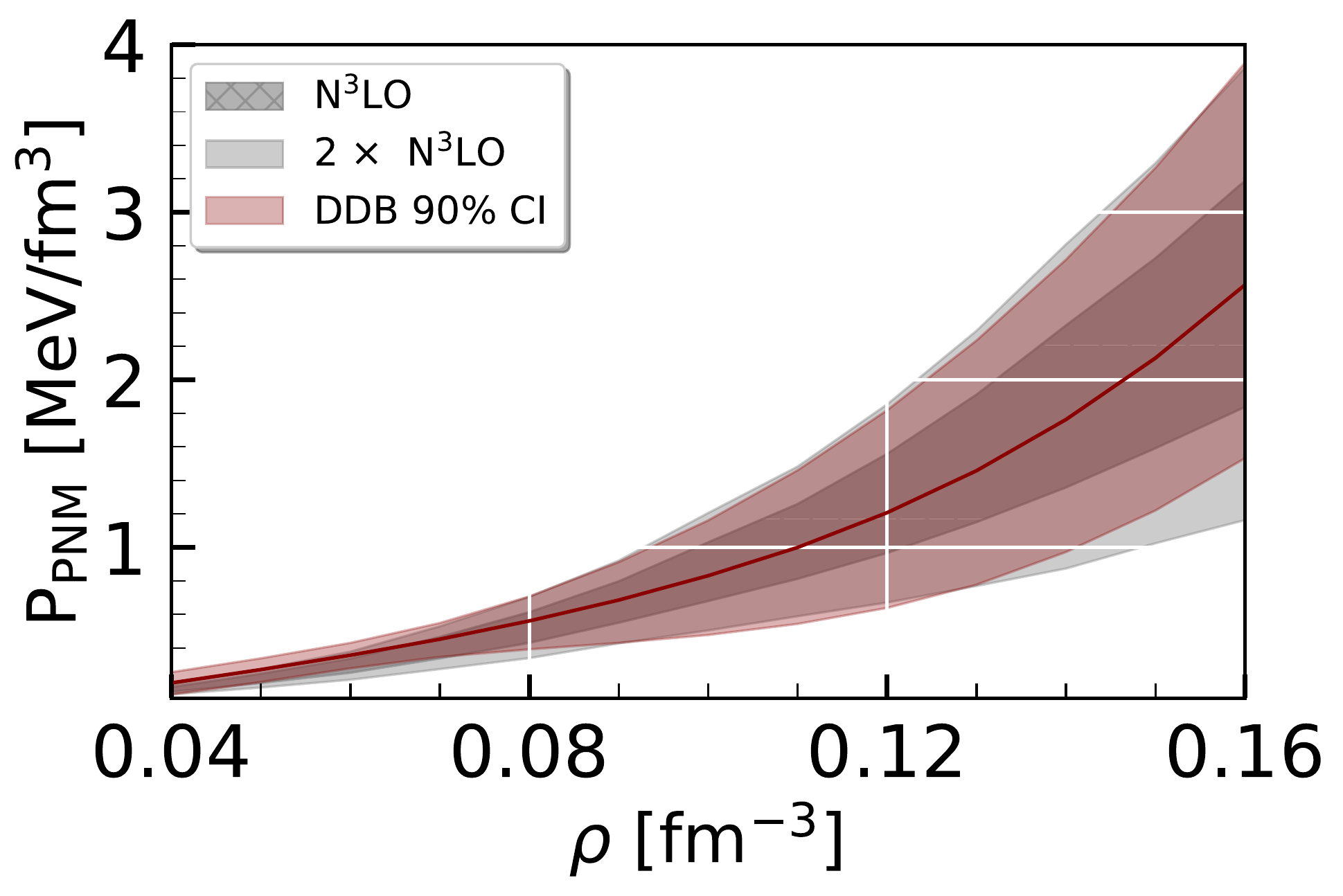}
\caption{ The pressure of low density neutron matter from a N$^3$LO calculation in $\chi$EFT
\cite{Hebeler2013}. The 90$\%$ CIs of the pressure of the low density neutron matter for DDB  is also compared.  It is to be noted that we consider $2\times$ N$^3$LO uncertainty and \blu{three intermediate points} in the likelihoods of the present calculation \label{T:fig2}.}
\end{figure}
{In Fig. \ref{T:fig2} we plot the low-density EOS for PNMs with 90\% CI for DDB. The results are obtained from the posterior distributions of the DDH parameters {corresponding to}  DDB  as listed in Table \ref{tab3}. For comparison, we also show the low density pressure band for pure neutron matter from $\chi$EFT. {The 90\% CI of low-density PNM pressure obtained for the constrained DDB model is in good agreement with these results:} it overlaps mostly with the two times $\chi$EFT band. The low density pressure constraints for PNM  plays a key role to constrain the density dependence of the symmetry energy and, thus, NS properties {at low mass}.}

{The Fig. \ref{M:fig1} shows the 90\% CIs for the pressure of $\beta$-equilibrium NS matter obtained from the posterior distributions of the DDH parameters of DDB  (dark red band) set. {For comparison
we also plot the constraints for $\beta$-equilibrium NS matter EOS obtained from GW170817 analysis \cite{LIGOScientific:2018cki}. The 90\% CI of  $\beta$-equilibrium  pressure as a function of baryon density obtained for DDB is fully compatible with the GW170817 constraints. The analysis performed for GW170817 did not impose the 2$M_\odot$ constraint.
}}

\begin{figure}[h]
\includegraphics[width=0.45\textwidth]{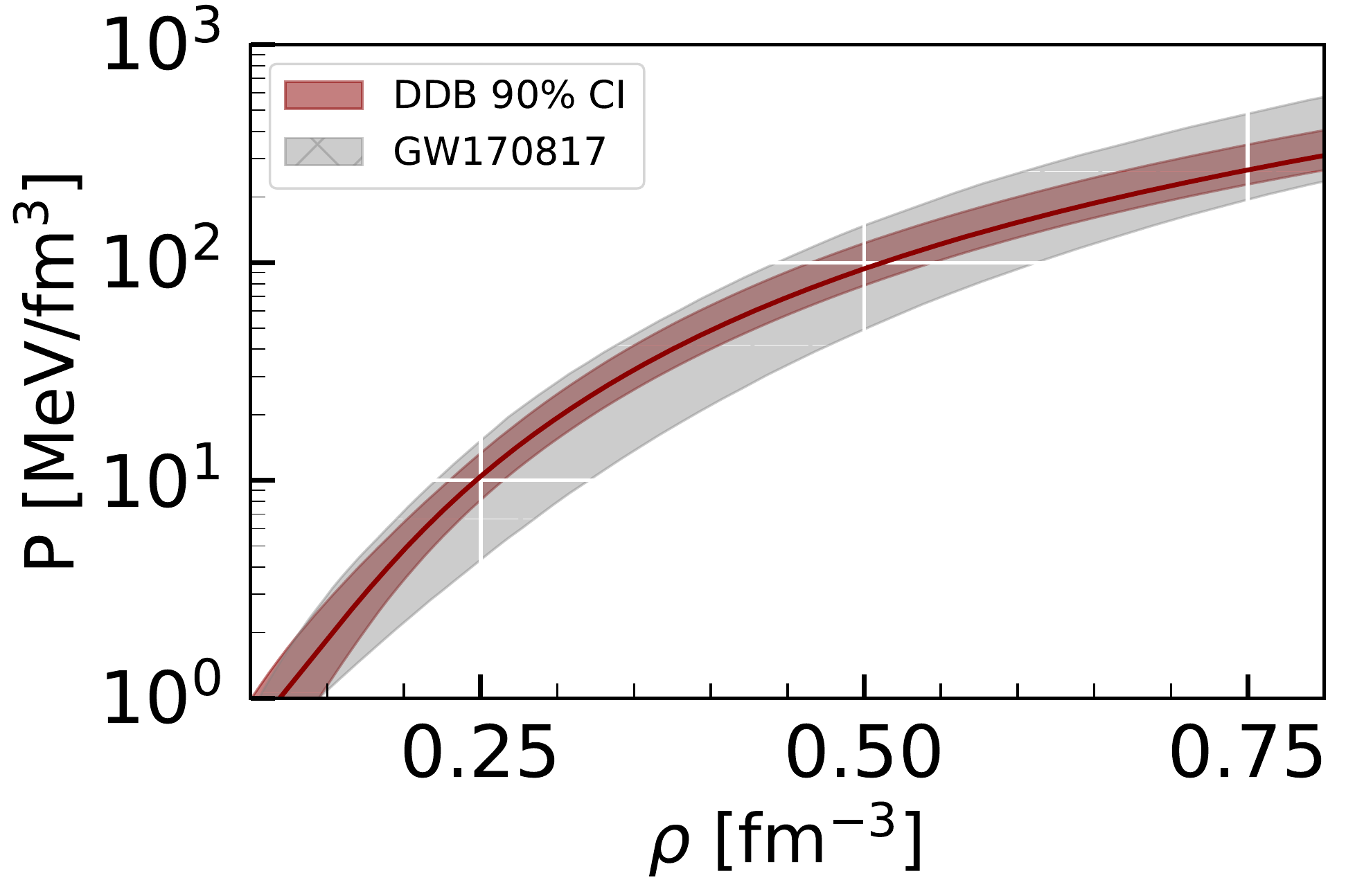}
\caption{The 90\% CIs for the pressure of NS matter as a function of the baryon density $\rho$ for DDB (dark red). For comparison we also plot the constraints for pressure obtained in GW170817. 
\label{M:fig1}} 
\end{figure}
\begin{table*}[hbt!]
\centering
\caption{The median values of DDH model parameters, namely $\Gamma_{\sigma,0}$, $\Gamma_{\omega,0}$, $\Gamma_{\varrho,0}$, $a_{\sigma}$, $a_{\omega}$ and $a_{\varrho}$ along with 68\%(90\%) CI obtained for DDB set using prior sets P defined in Table \ref{tab2}. The nucleon, $\omega$ meson, $\sigma$ meson and $\varrho$ meson masses are 939, 783, 550 and 763 MeV, respectively. \label{tab3}}
 \setlength{\tabcolsep}{5.5pt}
      \renewcommand{\arraystretch}{1.8}
\begin{tabular}{ccccccc}
\toprule
    & $\Gamma_{\sigma,0}$                  & $\Gamma_{\omega,0}$                   & $\Gamma_{\varrho,0}$               & $a_{\sigma}$                  & $a_{\omega}$                   & $a_{\varrho}$                  \\ \hline
DDB & $8.983_{-0.541 (0.802)}^{+0.547(0.855)}$ &
$10.699_{-0.851 (1.273)}^{+0.833(1.299)}$ &
$3.974_{-0.229 (0.393)}^{+0.218(0.361)}$ &
$0.079_{-0.016 (0.025)}^{+0.019(0.030)}$ &
$0.038_{-0.027 (0.034)}^{+0.038(0.059)}$ &
$0.544_{-0.142 (0.255)}^{+0.116(0.185)}$ \\
 \toprule
\end{tabular}
\end{table*}
\begin{figure*}[hbt!]
\includegraphics[width=0.95\textwidth]{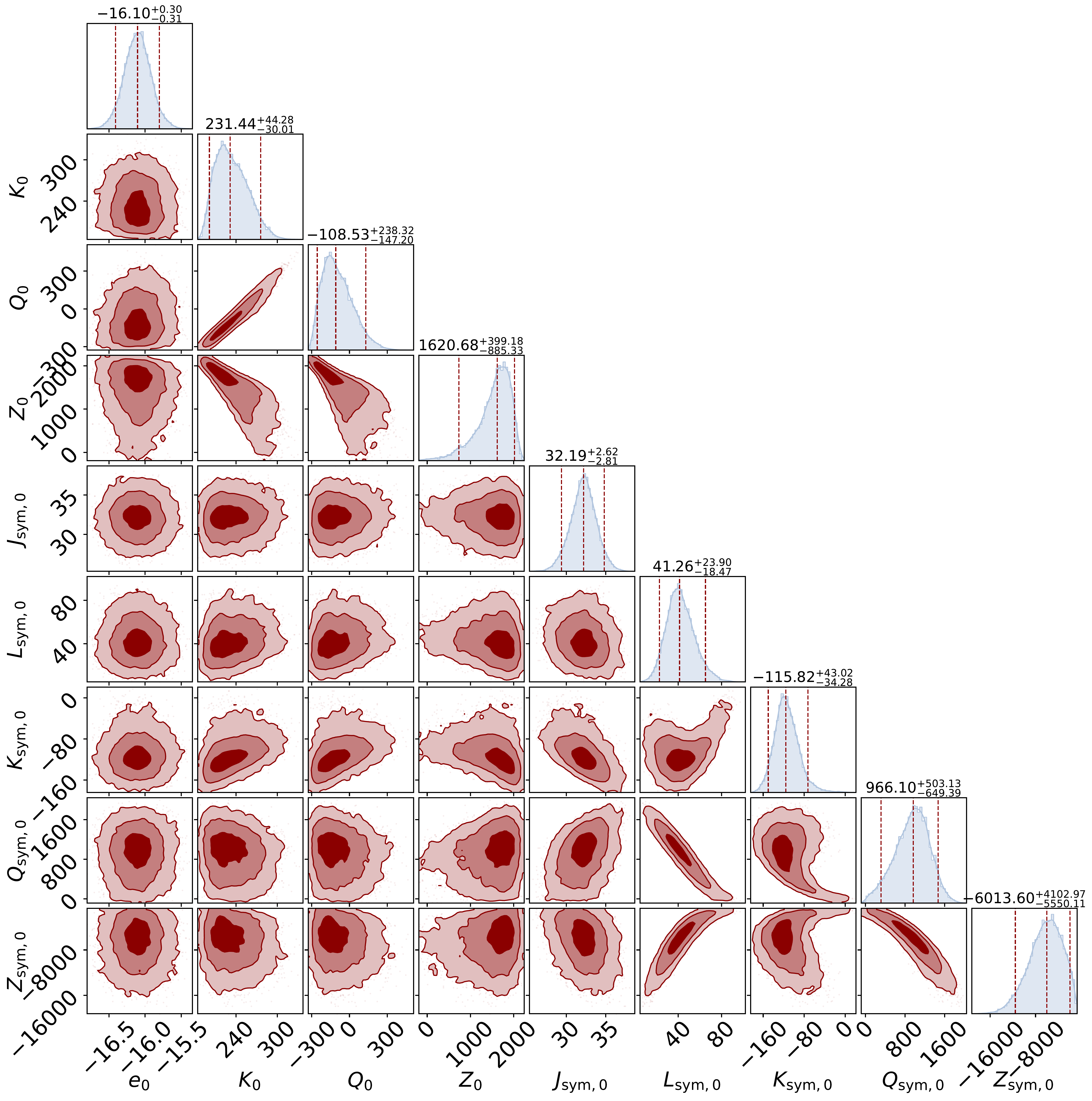}
\caption{Corner plots for the marginalized posterior distributions of the NMPs (MeV) obtained from  DDB (dark red) set of EOS for the neutron star matter, Eqs. (\ref{x0}) and (\ref{xsym}) \label{T:fig3}. The vertical lines indicate 90\% min, median and 90\% max CI, respectively,  and the different tonalities from dark to light indicate, respectively, the 1$\sigma$, 2$\sigma$, and 3$\sigma$ CI.}
\end{figure*}

\begin{figure*}[hbt!]
\includegraphics[width=0.95\textwidth]{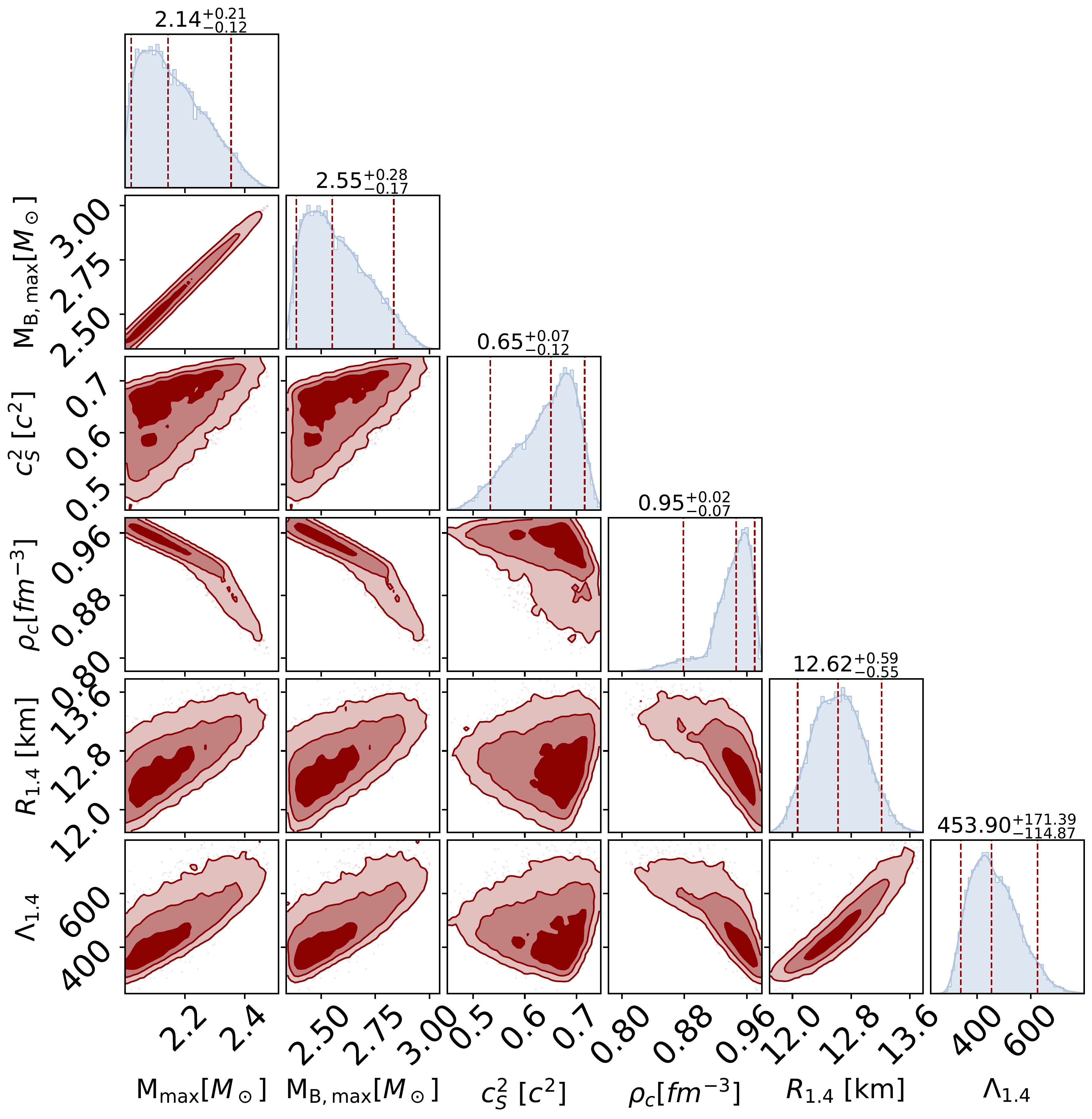}
\caption{Corner plots for the marginalized posterior distributions of neutron star properties, namely gravitational mass $M_{\rm max}$, baryonic mass $M_{\rm B, max}$, the square of central speed of sound $c_s^2$
, the central baryonic density $\rho_{c}$, the radius $R_{1.4}$ and  the dimensionless tidal deformability $\Lambda_{1.4}$ for 1.4 $M_\odot$ NS 
for the model DDB (dark red)  with prior set P defined in Table \ref{tab2}. The vertical lines indicate 90\% min, median and 90\% max CI, respectively, and the different tonalities from dark to light indicate, respectively, the 1$\sigma$, 2$\sigma$, and 3$\sigma$ CI.  \label{T:fig3b}}
\end{figure*}

{With the calculated posterior sets of DDH parameters, we perform a statistical analysis of the NMPs and neutron star properties, namely, mass, radius, central speed of sound and energy density,  and dimensionless tidal deformability.}
{In Table \ref{tab4}, we present the median values and the associated 50\% , 68 \% , 90\% and 95\% uncertainties of the NMPs and of some NS properties, namely, the following properties of the maximum mass star,
 the gravitational mass M$_{\rm max}$, the baryonic mass $M_{\rm B, max}$, the square of central speed of sound $c_s^2$, 
 the central energy density $\varepsilon_{c}$ and  the radius $R_{\rm max}$, 
 as well as the radius   and  the dimensionless tidal deformability for 1.4, 1.6, 1.8 and 2.08 $M_\odot$ NS obtained for the marginalized posterior distributions of the  DDH parameters. The NS masses and radii were calculated from the TOV equations \cite{TOV1,TOV2}  and the tidal deformability $\Lambda$ from the equations obtained in \cite{Hinderer2008}. In Figs. \ref{T:fig3} and \ref{T:fig3b} are given the corner plots for the  same quantities, respectively,  NMPs  and NS properties.}

In Fig. \ref{M:fig1a} we plot (left) the pressure for symmetric nuclear matter $P_{\rm SNM}$  and (right) symmetry energy ($S(\rho)$) as a function of number density together with  90\% CI for DDB set.
In the left panel we also include for reference the constraint obtained from heavy ion collision flow data on the pressure of symmetry nuclear matter \cite{Danielewicz2002pu}, which,  however, is not totally model independent. 
Let us recall that the nuclear model used to analyse the experimental data in \cite{Danielewicz2002pu}  does not predict two solar masses \cite{Constantinou2015} and, therefore, it is not surprising that the set DDB contains a large set of stiffer EOS. The median value of set DDB essentially coincides with the upper limit of the HIC constraint in the intermediate density region, and, therefore, more than 50\% of the EOS are out of the HIC predicted region.  {For comparison, in the right panel we also plot the constraints on symmetry energy obtained in nuclear structure studies involving excitation energies to isobaric analog states (IAS) \cite{Danielewicz:2013upa} and our result is in good agreement with them.} 
The $\chi$EFT PNM EOS constraints affect quite strongly the density dependence of symmetry energy: for instance, the slope of the symmetry energy, $L_{\rm sym,0}$ is concentrated between $\approx 30$ and 55 MeV, (68\% CI) although we may have values as high as $\approx$70 MeV. 

\begin{figure*}[t]
\centering
\includegraphics[width=0.44\textwidth]{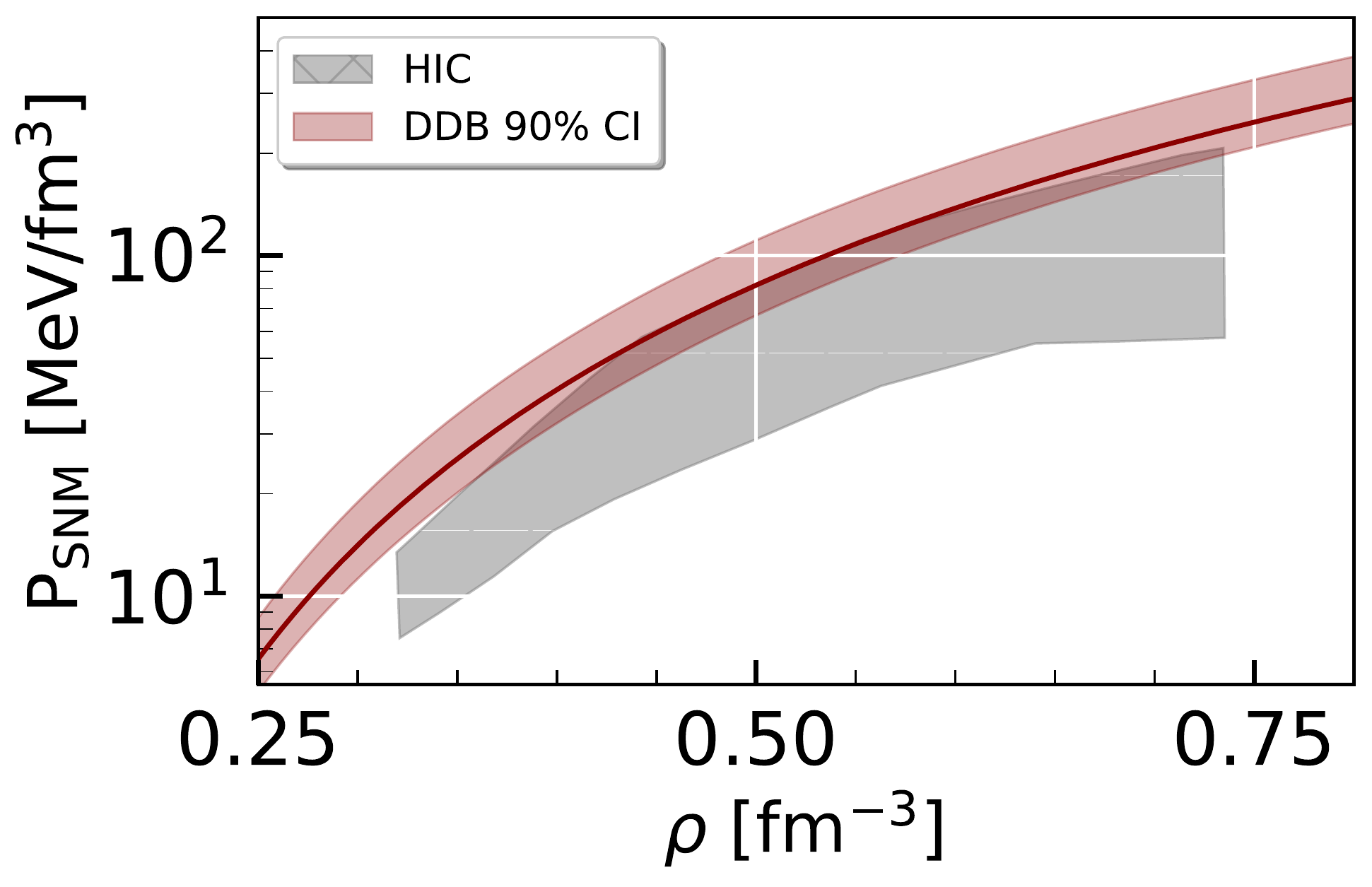}
\includegraphics[width=0.44\textwidth]{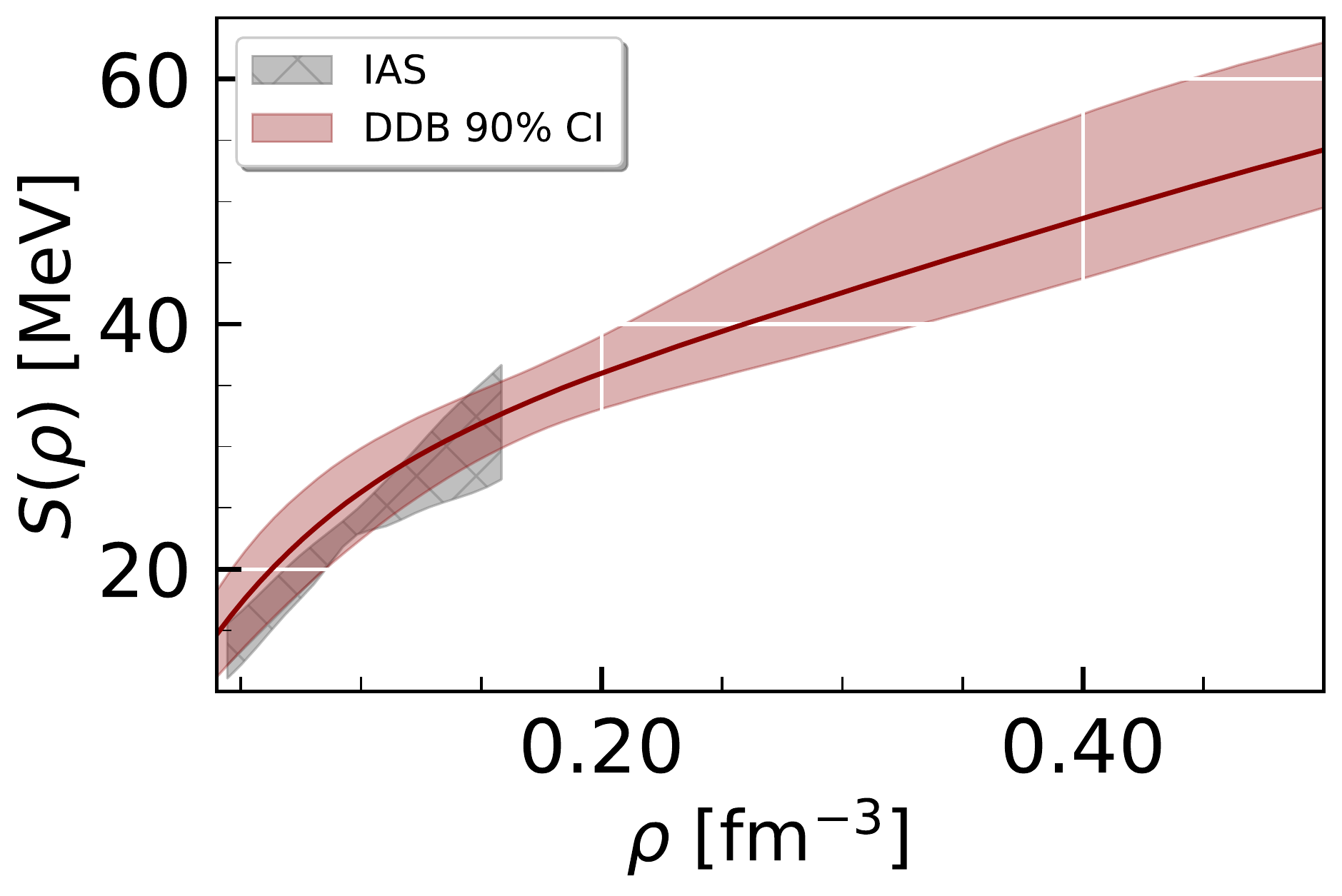}
\caption{(Left) The 90\% credible interval of the pressure for symmetric nuclear matter $P_{\rm SNM}$ and (right) symmetry energy $S(\rho)$ as a function of number density, together with the median (full lines) for the DDB  (dark red) sets. {In the left panel  heavy ion collision flow data for symmetry nuclear matter (gray band) have also been included \cite{Danielewicz2002pu}.} {The constraints on the symmetry energy from IAS \cite{Danielewicz:2013upa} are also displayed in the right panel.}}
\label{M:fig1a}
\end{figure*}
{In  \cite{Vidana2009}, it was found a linear correlation between $L_{\rm sym,0}$ and $K_{\rm sym,0}$ from a set of nuclear matter models based in Skyrme forces, a RMF approach or a microscopic approach. A similar correlation was determined in \cite{Tews:2016jhi} just from Skyrme forces, see the review \cite{Li:2019xxz} also discussing the role of $K_{\rm sym,0}$ on the determination of the core-curst transition density. Our predictions for $K_{\rm sym,0}$ are compatible with the values expected from that correlation and the predicted $L_{\rm sym,0}$. Based on a Taylor expansion EOS, in \cite{Zhang:2019fog} NS mass constraints have been imposed to define the valid domain for the NMPs $Q_0-K_{\rm sym,0}-Q_{\rm sym,0}$. While our median value for  $Q_{\rm sym,0}$ is within the range obtained in this study, it lies  20\% above the \cite{Zhang:2019fog} upper limit, 800~MeV. Our 90\%CI value indicates that $Q_{\rm sym,0}$ can be as large as $\approx 1500$~MeV.
{Recently, a Bayesian analysis was done in the framework of a Taylor expansion  EOS with the prior informed through  LIGO/Virgo as well as NICER measurements to constrain the NMPs, see \cite{Thi:2021jhz}. Our NMPs associated with the symmetry energy are somewhat more constrained. However, the choice of likelihood for $\chi$EFT is very different in both approaches and this may justify the differences.}

Concerning the isoscalar skewness, in  \cite{Zhang:2019fog} it is shown that $Q_0$ is strongly constrained by the NS maximum mass and causality, and the  range $-200$ to 200 MeV has been identified as compatible with observations. In our study at 90\% (95\%) CI, we have determined
$-256$ MeV$<Q_0<130$ MeV ($-271$ MeV$<Q_0<183$ MeV) with similar observational constraints,  quite compatible with the range calculated in \cite{Zhang:2019fog}, although with a smaller lower limit. 

Two aspects may explain the different parameters determined in both studies: on one side our approach is casual from the beginning, and a second reason is the fact that in \cite{Zhang:2019fog} the higher order parameters should be interpreted as an effective one since the Taylor expansion EOS does not contain terms beyond the third order, as discussed in \cite{Tovar2021}.}

\begin{table*}
\caption{The median values and the associated 50\% , 68 \% , 90\% and 95\% uncertainties of the NMPs introduced in Sec. \ref{model}, Eq. (\ref{x0}) and (\ref{xsym}), and NS properties, the gravitational mass $M_{\rm max}$, baryonic mass  $M_{\rm B, max}$, radius $R_{\rm max}$, central energy density $\varepsilon_c$, central number density for baryon $\rho_c$ and square of central speed of sound $c_s^2$ of the maximum mass NS, as well as the radius   and  the dimensionless tidal deformability for 1.4, 1.6, 1.8 and 2.08 $M_\odot$ NS, and also the combined tidal deformability $\tilde \Lambda$ for the GW merger with $q=1$ obtained for the DDB  set using prior sets P defined in Table \ref{tab2}.  \label{tab4}}
 \setlength{\tabcolsep}{5.2pt}
      \renewcommand{\arraystretch}{1.4}
\centering
\begin{tabular}{cccccccccccc}
\toprule
\multicolumn{2}{c}{\multirow{3}{*}{Quantity}} & \multirow{3}{*}{Units} & \multirow{3}{*}{median} & \multicolumn{8}{c}{Confidence Interval (CI)}                                                                      \\ \cline{5-12} 
\multicolumn{2}{c}{}                          &                        &                         & \multicolumn{2}{c}{$50\%$} & \multicolumn{2}{c}{$68\%$} & \multicolumn{2}{c}{$90\%$} & \multicolumn{2}{c}{$95\%$} \\ \cline{5-12} 
\multicolumn{2}{c}{}                          &                        &                         & min          & max         & min          & max         & min          & max         & min          & max         \\ \hline
\multirow{10}{*}{NMP} & $\rho_0$                 & fm$^{-3}$              & $0.153$                 & $0.150$      & $0.155$     & $0.149$      & $0.156$     & $0.147$      & $0.158$     & $0.147$      & $0.159$     \\
                      & $\varepsilon_0$          & \multirow{9}{*}{MeV}   & $-16.10$                & $-16.23$     & $-15.98$    & $-16.29$     & $-15.93$    & $-16.41$     & $-15.80$    & $-16.47$     & $-15.73$    \\
                      & $K_0$                    &                        & $231$                   & $216$        & $250$       & $210$        & $259$       & $201$        & $276$       & $198$        & $285$       \\
                      & $Q_0$                    &                        & $-109$                  & $-182$       & $-16$       & $-211$       & $33$        & $-256$       & $130$       & $-271$       & $183$       \\
                      & $Z_0$                    &                        & $1621$                  & $1340$       & $1826$      & $1159$       & $1902$      & $735$        & $2020$      & $531$        & $2066$      \\
                      & $J_{\rm sym,0}$          &                        & $32.19$                 & $31.11$      & $33.22$     & $30.56$      & $33.73$     & $29.38$      & $34.81$     & $28.80$      & $35.36$     \\
                      & $L_{\rm sym,0}$          &                        & $41.26$                 & $33.50$      & $50.26$     & $29.82$      & $54.90$     & $22.79$      & $65.16$     & $19.53$      & $70.64$     \\
                      & $K_{\rm sym,0}$          &                        & $-116$                  & $-130$       & $-100$      & $-137$       & $-92$       & $-150$       & $-73$       & $-157$       & $-60$       \\
                      & $Q_{\rm sym,0}$          &                        & $966$                   & $710$        & $1186$      & $583$        & $1277$      & $317$        & $1469$      & $202$        & $1567$      \\
                      & $Z_{\rm sym,0}$          &                        & $-6014$                 & $-8156$      & $-4043$     & $-9234$      & $-3232$     & $-11564$     & $-1911$     & $-12617$     & $-1488$     \\
                      &                          &                        &                         &              &             &              &             &              &             &              &             \\
\multirow{15}{*}{NS}  & $M_{\rm max}$            & M $_\odot$             & $2.144$                 & $2.076$      & $2.234$     & $2.052$      & $2.277$     & $2.021$      & $2.355$     & $2.011$      & $2.383$     \\
                      & $M_{\rm B, max}$         & M $_\odot$             & $2.552$                 & $2.461$      & $2.672$     & $2.428$      & $2.731$     & $2.386$      & $2.835$     & $2.375$      & $2.875$     \\
                      & $c_{s}^2$                & $c^2$                  & $0.65$                  & $0.60$       & $0.69$      & $0.58$       & $0.70$      & $0.53$       & $0.72$      & $0.52$       & $0.72$      \\
                      & $\rho_c$                 & fm$^{-3}$              & $0.946$                 & $0.929$      & $0.959$     & $0.919$      & $0.963$     & $0.879$      & $0.970$     & $0.862$      & $0.972$     \\
                      & $\varepsilon_{c}$        & MeV fm$^{-3}$          & $1282$                  & $1211$       & $1348$      & $1180$       & $1375$      & $1122$       & $1426$      & $1101$       & $1444$      \\
                      & $R_{\rm max}$            & \multirow{5}{*}{km}    & $11.09$                 & $10.84$      & $11.37$     & $10.74$      & $11.49$     & $10.56$      & $11.74$     & $10.50$      & $11.84$     \\
                      & $R_{1.4}$                &                        & $12.62$                 & $12.37$      & $12.87$     & $12.27$      & $12.98$     & $12.07$      & $13.21$     & $11.99$      & $13.30$     \\
                      & $R_{1.6}$                &                        & $12.53$                 & $12.27$      & $12.81$     & $12.15$      & $12.93$     & $11.95$      & $13.18$     & $11.87$      & $13.28$     \\
                      & $R_{1.8}$                &                        & $12.36$                 & $12.06$      & $12.69$     & $11.94$      & $12.83$     & $11.72$      & $13.11$     & $11.64$      & $13.21$     \\
                      & $R_{2.08}$              &                        & $12.01$                 & $11.59$      & $12.43$     & $11.41$      & $12.60$     & $11.10$      & $12.93$     & $10.99$      & $13.04$     \\
                      & $\Lambda_{1.4}$          & \multirow{5}{*}{-}     & $454$                   & $398$        & $523$       & $375$        & $555$       & $339$        & $625$       & $326$        & $655$       \\
                      & $\Lambda_{1.6}$          &                        & $185$                   & $158$        & $218$       & $148$        & $234$       & $132$        & $269$       & $126$        & $283$       \\
                      & $\Lambda_{1.8}$          &                        & $79$                    & $65$         & $97$        & $60$         & $106$       & $52$         & $125$       & $49$         & $133$       \\
                      & $\Lambda_{2.08}$        &                        & $22$                    & $16$         & $31$        & $14$         & $34$        & $11$         & $43$        & $10$         & $46$        \\
                      & $\tilde \Lambda_{q=1.0}$ &                        & $529$                   & $465$        & $608$       & $439$        & $645$       & $398$        & $724$       & $382$        & $759$    \\  \hline
\end{tabular}
\end{table*}
Concerning the NS properties, we conclude that: (i) the NS maximum mass is predicted in the range from  2.052-2.277 (2.021-2.355) $M_\odot$ for DDB  in 68\%  (90\%) CI, with the 95\% CI extreme 2.383 $M_\odot$ for DDB set. \blu{Note that this value
is just slightly smaller than the DD2 and DDME2 maximum mass, respectively, 2.42 $M_\odot$ \cite{Typel2009} and 2.48 $M_\odot$ \cite{Fortin2016}. Outside the 95\% CI, we find  DDB EOS that also describe stars with $M\sim 2.5M_\odot$ but these will be discussed later}; (ii) the square of speed of sound at the center of the maximum star is strongly constrained 
and the value is in the range 0.58-0.70 $c^2$ (0.53-0.72 $c^2$) at  68\%(90\%) CI.  {Having undertaken a causal approach it is interesting to notice that the speed of sound in maximum mass stars is still far from $c$}; (iii) the  central energy density of the maximum mass star for the DDB  set is  about 10\%  smaller than the value obtained not imposing the two solar mass constraint, signaling a stiffer EOS, e.g. less compressible; (iv) the radius and tidal deformability are quantities that are also strongly affected by the two solar mass constraint: for the DDB  set the minimum radius is $\approx$0.5-1~km larger and the minimum tidal deformability $\approx 150-230$ larger .   For a 1.4 $M_\odot$ star we get at 90\% CI $R_{1.4}\in [12.07,13.21]$ and $\Lambda_{1.4}\in [339,625]$. At 95\% CI, we do not get radii below $\sim 12$~km.}
The radius and dimensionless tidal deformability are in good agreement with NICER and GW170817, respectively, as seen in Fig. \ref{M:fig3} and discussed below. {Let us point out that the  prediction for the tidal deformability $\Lambda_{1.4}$ is coincident with the range of values predicted in \cite{LIGOScientific:2018cki} imposing no mass constraint as shown in Fig. \ref{T:f9}}. 

{We next discuss the lower bounds of the tidal deformability  of a 1.36~$M_\odot$ star, which would be the NS mass of each NS in the binary associated to the GW170817 if it would have been symmetric, i.e. $m_1=m_2$. In this case the effective $\tilde \Lambda=\Lambda(M=1.36\, M_\odot)$. It was shown by several authors that the follow up electromagnetic counterparts, the gamma-ray
burst GRB170817A \cite{LIGOScientific:2017zic}, and the electromagnetic
transient AT2017gfo \cite{LIGOScientific:2017ync}, set constraints on the lower limit of the effective tidal deformability $\tilde\Lambda$, in particular, in \cite{Radice:2017lry}  the lower limit  $\tilde\Lambda\gtrsim 300$ was obtained and in \cite{Kiuchi_2019} $\tilde\Lambda\gtrsim 242$. From our  set DDB, satisfying the two solar mass constraint,  we get $\Lambda_{1.36} >382$ at 95\% CI for the DDB set, slightly larger than the proposed limits.}
 
\begin{figure}
\includegraphics[width=0.48\textwidth]{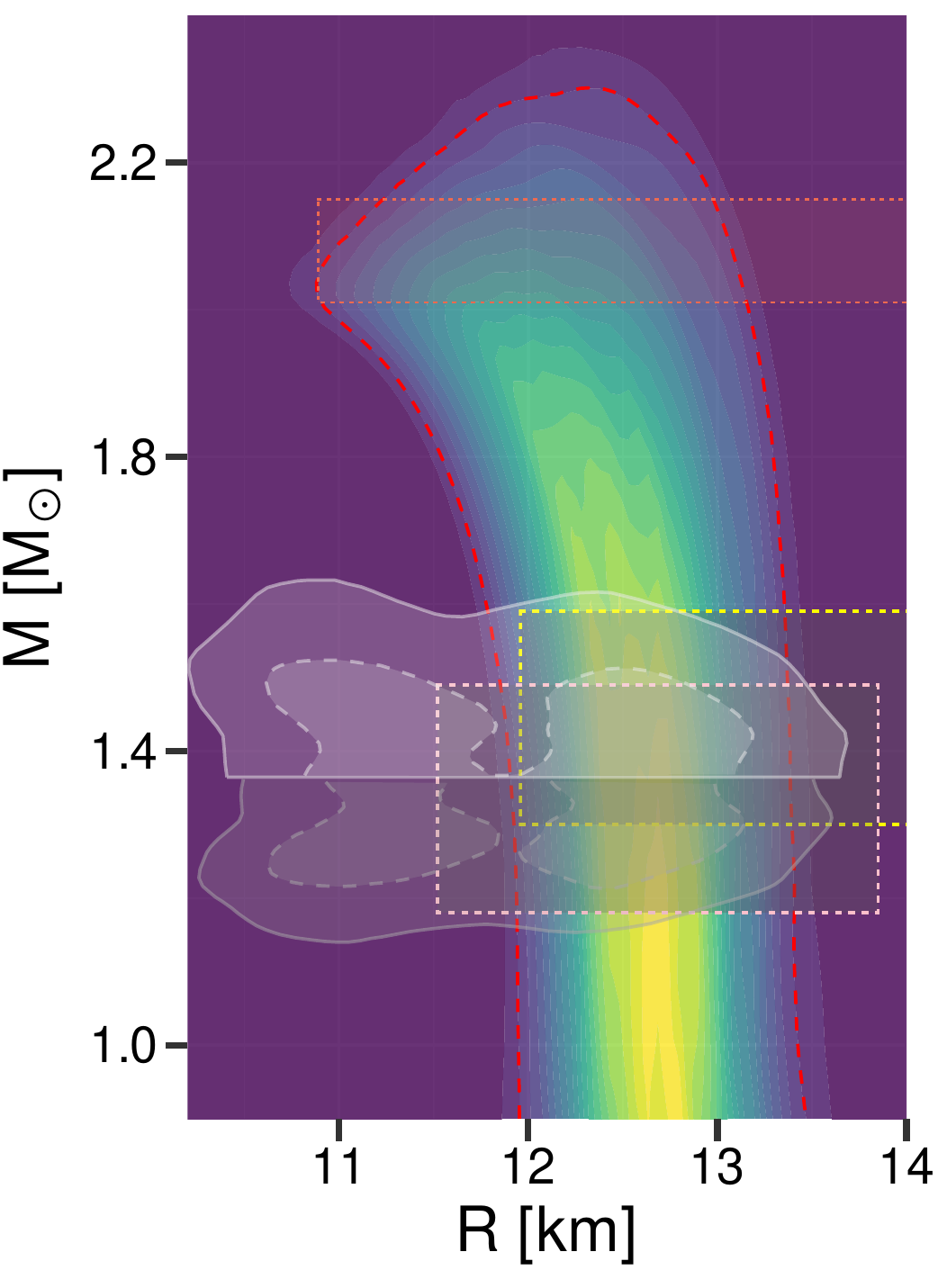}
\caption{Plot of the joint probability distribution  $P(M,R)$ for model  DDB. The colors levels are \crule[color1]{0.25cm}{0.25cm} (0-0.05), \crule[color2]{0.25cm}{0.25cm} (0.05-0.10), \crule[color3]{0.25cm}{0.25cm} (0.10-0.15),
$\cdots$, \crule[color18]{0.25cm}{0.25cm} (0.85-0.90), \crule[color19]{0.25cm}{0.25cm} (0.90-0.95), \crule[color20]{0.25cm}{0.25cm} (0.95-1.00). 
The red dashed line represents the 90\% CI. The top and bottom gray regions indicate, respectively, the 90\% (solid) and 50\% (dashed) CI of the LIGO/Virgo analysis for each binary component from the GW170817 event \cite{LIGOScientific:2018hze}. The rectangular regions enclosed by dotted lines indicate the constraints from the millisecond pulsar PSR J0030+0451 NICER x-ray data  \cite{Riley:2019yda,Miller:2019cac} and PSR J0740+6620 \cite{Miller:2021qha}.
\label{M:fig3}} 
\end{figure}

\begin{figure*}
\includegraphics[width=0.98\textwidth]{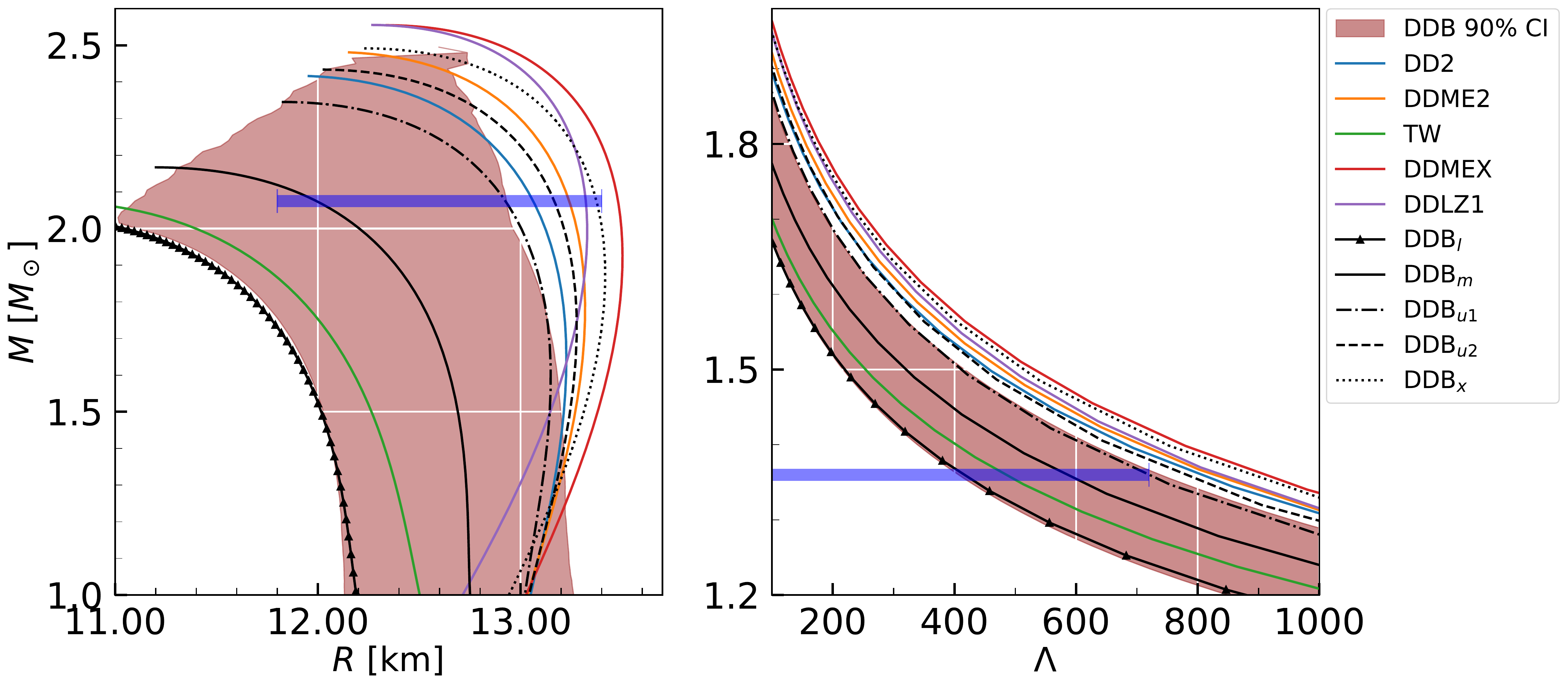}
\caption{The 90\% CI for the conditional probabilities $P(R|M)$ (left) and $P(\Lambda|M)$ (right) for DDB  (dark red). For reference on the left panel, results for  models TW \cite{Typel1999}, DD2 \cite{Typel2009}, DDME2 \cite{Lalazissis2005}, DDMEX \cite{Taninah:2019cku,Huang:2020cab}, DD-LZ1 \cite{Wei:2020kfb} and DDBl, DDBm, DDBu1, DDBu2 and DDBx are also shown. The blue horizontal bars indicate: on the left panel the 90\% CI radius for a 2.08$M_\odot$ star determined in \cite{Miller:2021qha} combining observational data from GW170817 and NICER as well as  nuclear data, on the right panel the 90\% CI obtained for the tidal deformability of a 1.36$M_\odot$ star in \cite{LIGOScientific:2018cki}. \label{M:fig2}} 
\end{figure*}
\begin{figure*}
\includegraphics[width=0.48\textwidth]{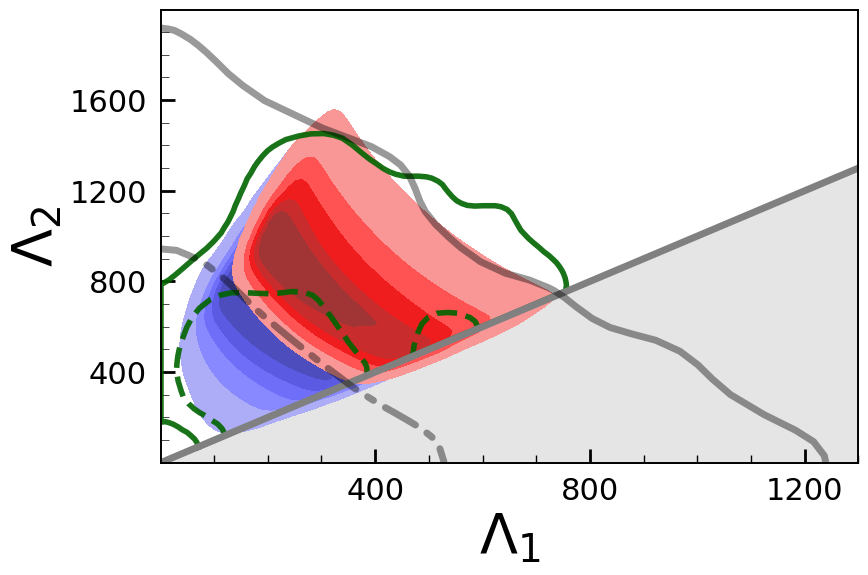}
\includegraphics[width=0.48\textwidth]{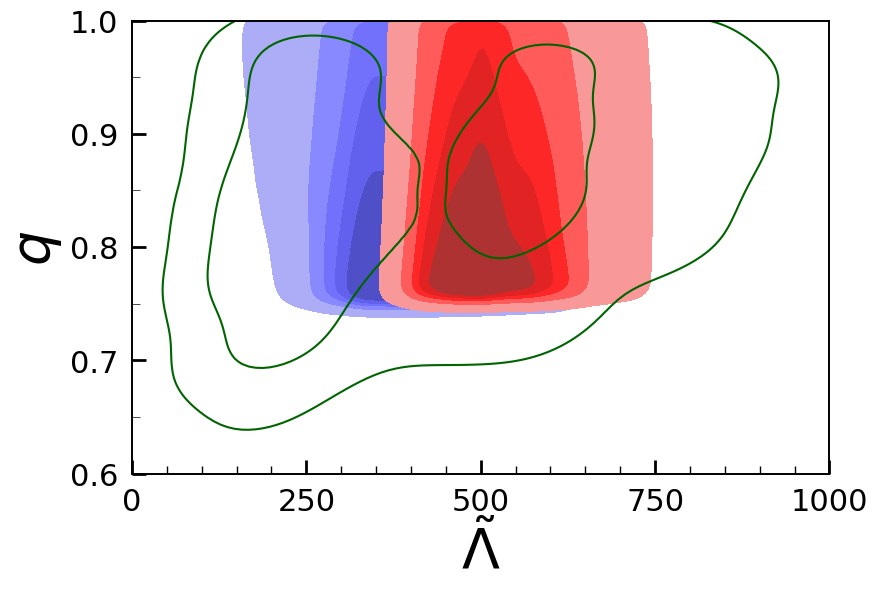}
\caption{(left) The probability distribution $P(\Lambda_1,\Lambda_2)$, where $\Lambda_1$ and $\Lambda_2$
are the dimensionless tidal deformability parameters of the binary neutron star merger from the GW170817 event, using the observed chirp mass of $M_{\rm chirp}=1.188$ M$_\odot$ and mass ratio $q=m_2/m_1$ ($0.7 <q< 1$), for the marginalized posterior distribution of DDH model parameters  with the prior set P, allowing, however, that the two solar mass constraint is not imposed (blue region). The gray solid (dashed) line represents the 90\%(50\%) CI from  the marginalized posterior for the tidal deformabilities of the two binary components of GW170817. The green solid (dashed) lines represent the 90\%(50\%) CI of the marginalized posterior for the tidal deformabilities of the two binary components of GW170817 using a parametrized EOS with a maximum mass requirement. (right) The 
probability distribution $P(q,\tilde\Lambda)$, the $\tilde \Lambda$ is the effective tidal deformability
in the binary with mass ratio $q$. The LIGO/Virgo Collaboration results \cite{LIGOScientific:2018hze} for the probability distribution function of the joint posterior is shown by the green color for 90\% CI and 50\% CI, respectively. The blue and red regions in both the panels represent samples not imposing and imposing  the two solar mass constraint, respectively. The last case corresponds to the DDB set.} 
\label{T:f9}
\end{figure*}

{In Fig. \ref{M:fig3}, we plot the joint PDs $P(M,R)$ of  the mass and the radius for DDB.
The red dashed line represents the 90\% CI. The color gradient from yellow to blue represents the highest to lowest probability. It can be seen that the probability $P(M,R)$ is highest for  a radius $\approx$12.7 km and a mass from 1 - 1.75 $M_\odot$. The lower bound of NS maximum mass with 68\% CI from marginalized PDs is 2.05 $M_\odot$ (see Table \ref{tab4}). So, below this lower bound {all masses have a similar} number of points and as we go to higher masses the number of points reduces as all the EOS correspond to PDs that have a different maximum mass. We also compare the 90\% CI of the $P(M,R)$ for DDB  set with those results obtained in GW170817 and NICER analysis (right). The upper {grey region limited by a solid (dashed) line} is the 90\% (50\%) CI of marginalized posterior for the mass M and  radius R obtained in the GW170817 analysis of the heaviest NS in the binary component using a parametrized EOS, where a lower limit on the maximum mass of 1.97 $M_\odot$ was imposed.  On the other hand the similar lower  shaded region represents the same but for the lighter mass in the binary of the GW170817 event. The rectangular regions enclosed by dotted lines indicate the constraints from the analysis of the millisecond pulsar PSR J0030+0451 NICER x-ray data  \cite{Riley:2019yda,Miller:2019cac}. It is to be noted that the 90\% CI in the NS mass and radius space for DDB, as represented by red dashed is in very good agreement with both GW170817 and NICER overlap region. The highest probability for the mass and radius calculated with the DDB  model lies precisely in the middle of the GW170817 and NICER overlap region. 
Considering a 1.4$M_\odot$ star,  
the dimensionless tidal deformability $\Lambda$  is 375-555 (326-655), and the radius is
12.27-12.98 (11.99-13.30) $\sim$ km at 68\% (95\%) CI.
The dimensionless tidal deformability for 1.4 $M_\odot$ NS predicted in GW170817 event with 90\% CI is below 780: this constraint is satisfied by DDB.  It should be pointed out, however, that the predicted value for the dimensionless tidal deformability  in GW170817 requires the specification of an EOS, and, therefore, is  model dependent. 
{We conclude that  the present NICER and GW170817 data cannot constrain further the uncertainties present in the DDB  model for the EOS, NMPs and  mass-radius region.} We expect that in the future, further strict constraints on joint PDs of mass and radius from either NICER or GW will reduce these uncertainties.}

\blu{ The above results have been obtained within two different methods, corresponding to around final selected  14000 EOSs in {\it Pymultinest} and 13000 EOSs in {\it Dynesty} which give very similar results. In order to understand which is the maximum mass described by our DDH model, we have looked for EOS that predict maximum masses above 2.48$M_\odot$. From the 225 EOS obtained most of them have a mass $\lesssim 2.5 M_\odot$ and an incompressibility of the order of 300 MeV. Parametrization DDBx plotted in Fig. \ref{M:fig2} is one of these EOS: it predicts a maximum mass of 2.5$M_\odot$ and has the following nuclear matter properties, $K_0=300$ MeV, $J_{sym,0}=30$ MeV and $L_{sym,0}=39$ MeV. As discussed below, part of the M-R curve lies outside the 90\% CI for the conditional probabilities $P(R|M)$.}

\blu{In Fig. \ref{M:fig2}, we plot the 90\% CI for the conditional probabilities $P(R|M)$ (left) and $P(\Lambda|M)$ (right) from the posterior distributions of the DDH parameters in  DDB  set (dark red shaded region).  This means that  from all the radii and tidal deformabilities obtained for a given mass, 90\% lie inside the interval represented. In this case the maximum mass corresponds precisely to the maximum mass inside the set DDB. }

\blu{For reference, we have also included 
in both panels of Fig. \ref{M:fig2}  several other DDH EOS known from the literature, in particular,  TW \cite{Typel1999}, DD2 \cite{Typel2009}, DDME2 \cite{Lalazissis2005},  DDMEX \cite{Taninah:2019cku,Huang:2020cab}, DD-LZ1 \cite{Wei:2020kfb} and five DDB models (DDBl, DDBm, DDBu1, DDBu2 and DDBx). The first five EOS are determined from models with density dependent couplings fitted to nuclear properties, and the last five 
 have been chosen from the set DDB and are given in the supplementary material.
 DDBl, DDBm, DDBu2 were chosen so that the radius of the 1.4$M_\odot$ star has the lower limit, a medium value and the upper limit  of the the 90\% CI for the conditional probabilities $P(R|M)$. We have also included DDBu1 that has a slightly lower $R_{1.4}$ than the upper limit but lies completely inside the 90\% CI for the conditional probabilities $P(R|M)$.}
\blu{
In the left panel we have also included an horizontal bar indicating the predicted radius of a 2.08$M_\odot$ star to 90\% credibility ($11.8 - 13.4$ km) as calculated in \cite{Miller:2021qha}  combining nuclear data and  observational data from GW170817 and NICER (from PSR J0030+0451 and PSR J0740+6620). This interval shrinks to $12.2 - 13.1$ km at 68\% credibility. The present  PSR J0704+6620 radius  determination by NICER   ignoring other information undertaken in \cite{Miller:2021qha,Riley:2021pdl} predicts a quite large interval, and does not  allow any conclusions to be drawn.} 

\blu{It is seen that several 
mass-radius curves lie partially outside the 90\% CI, DD2, DDME2, DDMEx, DDLZ1, DDBx and DDBu2. On the right panel, it is clear that these same models lie outside the 90\% CI obtained for the tidal deformability, even for low masses.  None of these models satisfies the constraint that GW170917 sets on the tidal deformability, $70<\tilde \Lambda=\Lambda(1.36 M_\odot)<720$, and indicated by the blue horizontal bar on the right panel.  }

\blu{What distinguishes the set of models plotted in  Fig. \ref{M:fig2} is the high density behavior of the EOS and a more precise determination of the radius of a two solar mass star will allow to distinguish between them. In fact, the density dependence of  the couplings allows for quite different behaviors in the high density range. The harder EOS are the ones that predict the larger masses. In common,  we see that many of these M-R curves present a back-bending behavior. If the radius of the canonical star with 1.4$M_\odot$ and a two solar mass NS are determined with a small enough uncertainty the different models may  be filtered. The 90\% credibility  radius of a 2.08$M_\odot$ star indicated by the horizontal bar does not exclude a region of the M-R diagram that within our model lies outside the the 90\% CI.
}

\blu{In  \cite{Wei:2020kfb,Taninah:2019cku} the models DD-MEX and DD-LZ1 predicting a 2.55$M_\odot$ maximum  mass and having, simultaneously, reasonable saturation nuclear matter properties, have been proposed, see the discussion in \cite{Huang:2020cab}. These models are based in the same framework as DD2 and DDME2. This parametrization seems to offer more freedom than the one  proposed in the present work, {allowing for a harder EOS at high densities and predicting larger radii}.   Our parametrization does not allow for  masses above $\approx$2.5$M_\odot$. In the future a more careful investigation of the possible density behavior of the couplings  and consequences will be carried out.}

\blu{Other models have predicted masses above 2.44$M_\odot$: (i) NL3$\omega\rho$ with $L_{\rm sym,0}$=55 MeV predicts a maximum mass of 2.75 $M_\odot$, but this EOS has a very hard isoscalar EOS, in particular, $K_0=271$MeV \cite{Fortin2016} and $\Lambda_{1.4}=1040$; (ii) BigApple \cite{Fattoyev:2020cws} describes a 2.6$M_\odot$ NS but does not satisfy PNM $\chi$EFT constraints; (iii) using a Taylor expansion EOS to describe nuclear matter maximum masses as high as 2.66$M_\odot$ were obtained. However, this is a non-relativistic approach  and it is necessary to filter the models that do not satisfy $c_s<1$, precisely the condition that defines the maximum mass upper limits.} 

{In Fig. \ref{T:f9} (left), we show the probability distribution of the dimensionless tidal deformability parameters $\Lambda_1$ and $\Lambda_2$ as $P(\Lambda_1,\Lambda_2)$ for the 2 objects involved in the BNS event from GW170817, with masses $m_1$ and $m_2$, using the observed chirp mass of $M_{\rm chirp}=1.188$ M$_\odot$ and mass ratio $q=m_2/m_1$ ($0.7 <q< 1$), for the marginalized posterior distribution of DDH model parameters of two cases: (i) a set similar to DDB but without having NS maximum mass constraints (blue) and (ii) for DDB set (red). For each EOS, we  obtain a curve in the $\Lambda_1$ and $\Lambda_2$ plane by varying $m_1$ in the range $1.36 < m_1 < 1.6$ M$_\odot$, and calculating  $m_2$  by keeping the chirp mass fixed at $M_{\rm chirp}=1.188$ M$_\odot$, as observed in the GW170817 event. 
We also show the constraints from GW170817 for comparison. The black solid (dashed) line represents the 90\%(50\%) CI from the marginalized posterior for the tidal deformabilities of the two binary components of GW170817. The green solid (dashed) lines represent the 90\%(50\%) CI of the marginalized posterior for the tidal deformabilities of the two binary components of GW170817 using a parametrized EOS with a maximum mass requirement of at least 1.97 $M_\odot$. In the right panel,  we present the PDs for $P(q,\tilde\Lambda)$ in the mass ratio $q$ and combined tidal deformability $\tilde \Lambda$ for the GW merger. The green lines are the 50\%(90\%) CI given by LIGO/Virgo analysis \cite{LIGOScientific:2018hze}. We see that both the $P(\Lambda_1,\Lambda_2)$ and $P(q,\tilde\Lambda)$ obtained with DDH parameters are in very good agreement with the GW170817 LIGO/Virgo results.}

{NICER has measured the equatorial circumferential radius of one of the highest mass ($2.072^{+0.067}_{-0.066}$ $M_\odot$)  pulsar PSR J0740 + 6620. This measurement of radius with 68\% CI is $12.39_{-0.98}^{+1.30}$ \cite{Riley:2021pdl}. We also investigate the prediction for the radius of a 2.08 $M_\odot$ NS within the DDB  set:  we have determined for the radius median value  $\approx$ 12.01 km, and for the 90\% CI
 $\approx$ 11.1 - 12.9 km.}

\begin{figure}[t]
\includegraphics[width=0.45\textwidth]{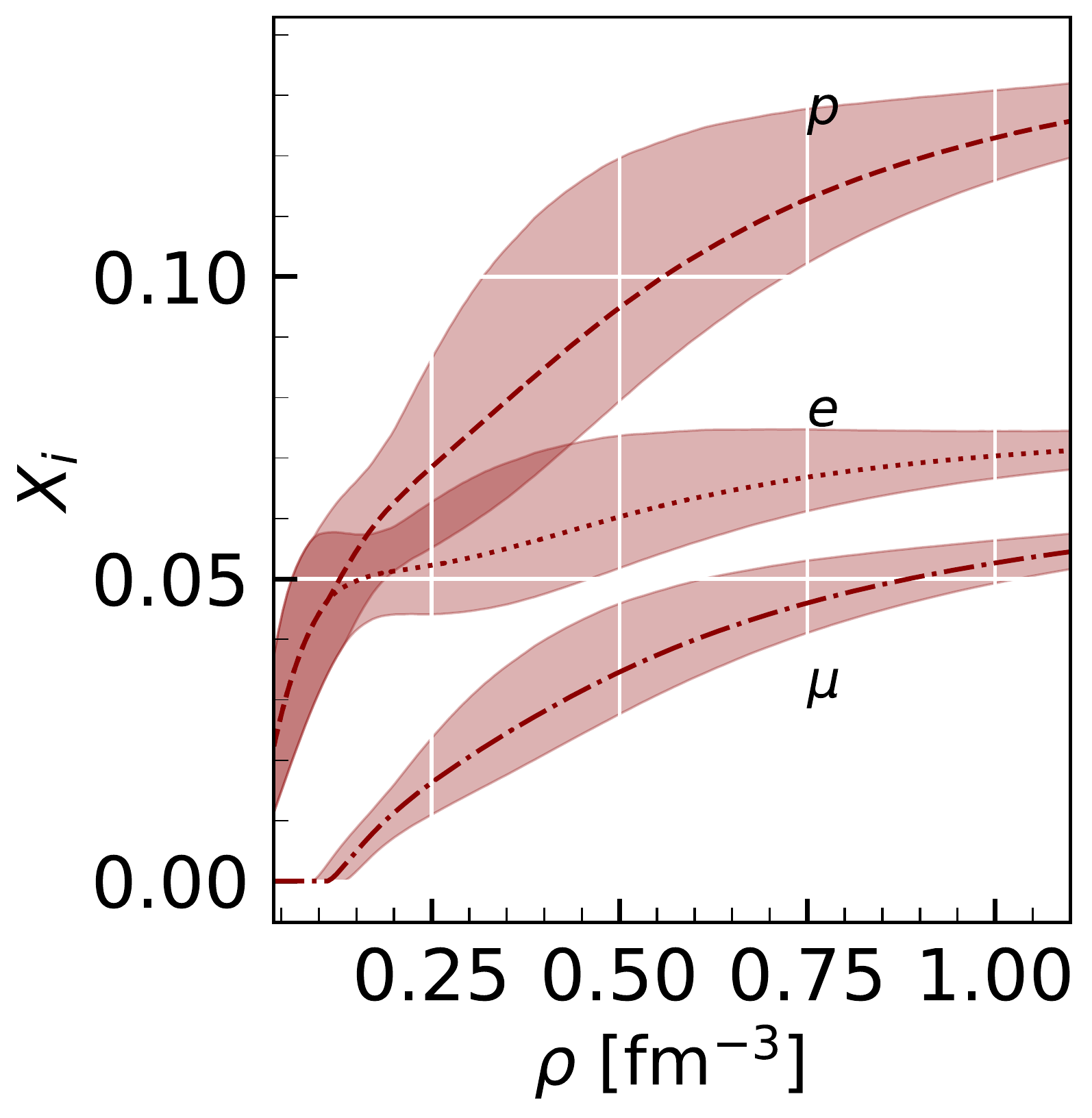}
\caption{The plot for the particle fraction $X_i$, for protons (p), electrons (e) and muons ($\mu$), along with 90\% CI as a function of baryon density $\rho$ for DDB (dark red). } 
\label{M:f}
\end{figure}
{The composition of NS may be constrained by cooling information: as soon as the nucleonic direct Urca process sets in the NS undergoes a super-fast cooling \cite{Lattimer:1991ib,Yakovlev:2000jp}. This is possible if the proton fraction attains the minimum threshold that, if muons are excluded, corresponds to 1/9 \cite{Lattimer:1991ib}.  Including muons increases this fraction to above 0.14 after muon onset, in \cite{Thi:2021jhz} a threshold of 0.135 and 0.138 was obtained, respectively for 1.4 and 2.0 $M_\odot$ stars. In Fig. \ref{M:f}, the proton, electron and muon fractions are plotted as a function of density. NS central densities in our sets lie below 1.1 fm$^{-3}$. We have verified that the present set of models does not predict nucleonic direct Urca  inside NS. A more careful analysis, also considering the opening of hyperonic Urca processes will be studied in the future. This agrees with conclusions drawn in \cite{Fortin2016,Fortin:2020qin,Fortin:2021umb} for DDH models such as DD2 and DDME2.}

\begin{figure}
\includegraphics[width=0.45\textwidth]{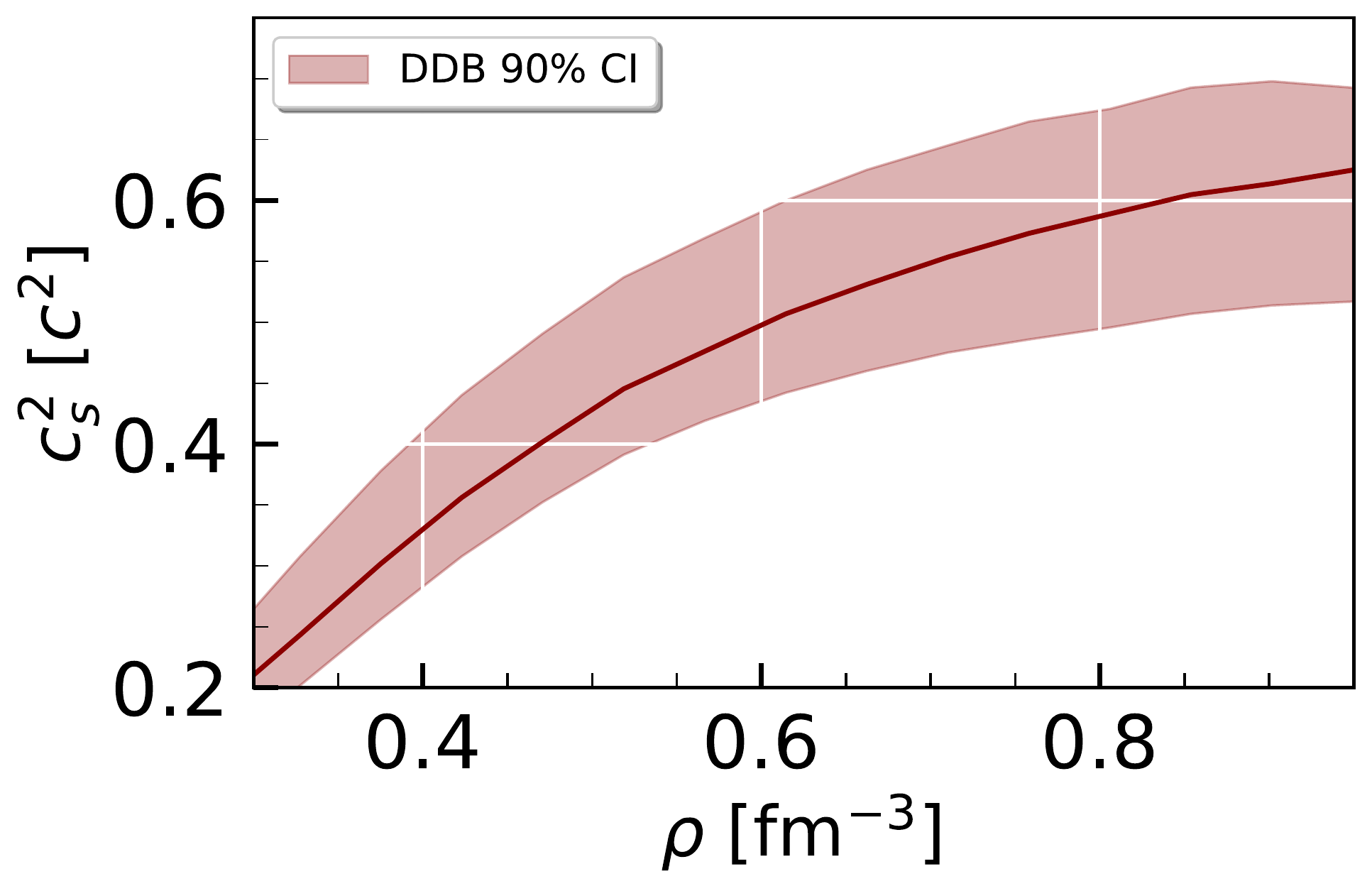}
\caption{The square of sound velocity $c_s^2$ with 90\% CI as a function of baryon density for DDB set (dark red).} 
\label{T:cs}
\end{figure}

\blu{Finally, let us also refer to the behavior of the sound velocity of the models within our DDB set, see Fig. \ref{T:cs}. As discussed in other works \cite{Bedaque:2014sqa,Alford:2013aca,Moustakidis:2016sab,Tews:2018kmu,Reed:2019ezm}  2$M_\odot$ NSs require a speed of sound well above the conformal limit $1/\sqrt{3}$,  implying that neutron star matter is a strongly  interacting system. This indicates that in order for neutron star matter to be able to counterbalance the gravitational attraction the energy density should increase slowly enough compared with the pressure increase, resulting in a large speed of sound. In the center of the NSs  the square of speed-of-sound squared is, in average, 0.65 but taking the 95\% CI  $0.51<c_s^2< 0.73$. As referred before, the present description of hadronic matter automatically limits the speed-of-sound to values below 1. In \cite{Ferreira:2021osk}, a larger central speed-of-sound was necessary to attain 2$M_\odot$ hybrid stars,  always above $\sqrt{0.7}$. As explained in \cite{Alford:2013aca}, this is necessary so that the quark core is able to support the nuclear mantle.} 

\section{Conclusions \label{con} }
Within a Bayesian approach, we have generated a set of models based on the RMF framework with density dependent coupling parameters and no non-linear mesonic terms. This set was constrained by the neutron matter $\chi$EFT EOS and  four saturation properties of nuclear matter: the saturation density, binding energy per particle and incompressibility and the symmetry energy. For the last  property an interval compatible with $\chi$EFT calculations was considered. Besides, the set was also constrained by imposing that the maximum star mass should  be at least 2$M_\odot$. 
It was verified that  the 90\% CI for the low density pure neutron matter calculated within the DDB set is  compatible with the one obtained from a precise N$^{3}$LO calculation in $\chi$EFT \cite{Hebeler2013}.

The main objective of the study is the determination of the domain of nucleonic neutron star EOS {based on a relativistic approach with minimal constraints.} In particular, we have analysed the behavior of the density dependence of the symmetry energy, the high density behavior of the EOS and the upper and lower limits for several NS properties. {We have verified that the posterior distribution of NS maximum mass, radii and tidal deformabilities are compatible with recent NS observables.}

The recent determination of the neutron skin thickness of $^{208}$Pb through PREX-II measurements \cite{PREX2}, $\Delta R_{\rm skin}=0.283\pm0.071$fm,  seems to indicate that the slope of the symmetry energy could be rather high,  $L_{\rm sym,0}=(106\pm37)$~MeV according to \cite{PREX2}, showing some tension with the results we have obtained. Other studies, however, have obtained smaller values for the slope.  In \cite{Essick2021}, undertaking an analysis that combines the  astrophysical data with constraints from PREX-II  and  $\chi$EFT calculations,  the authors have determined a   $^{208}$Pb neutron skin thickness equal to $0.17\pm0.04$~fm and a symmetry energy slope $L_{\rm sym,0}=53^{+14}_{-15}$~MeV. This last prediction for the slope
is compatible with the range of values determined with the DDB set. 

{The present study has enabled us to understand which are the limitations of the Taylor expansion EOS approach to determine the acceptable range of {values for higher order NMPs as the skewness $Q_0$ for the symmetric nuclear matter and the incompressibility $K_{\rm sym,0}$ and  skewness $Q_{\rm sym,0}$ for symmetry energy. Within a 90\% CI these last three quantities take the values $-256<Q_0<130$~MeV, $-150<K_{\rm sym,0}<-73$~MeV  and  $317<Q_{\rm sym,0}<1469$~MeV. } 
In the Taylor expansion approach, the isoscalar NMPs are constrained by causality conditions not intrinsic to the model. Besides, they should be interpreted as effective parameters since they have to describe effects of the missing higher terms \cite{Tovar2021}.}

NS properties have been studied and compared with recent observations, masses of pulsars PSR~J1614-2230  \cite{Demorest2010,Fonseca2016,Arzoumanian2017}, PSR~J0348 - 0432 \cite{Antoniadis2013},  PSR~J0740+6620  \cite{Fonseca:2021wxt} and very recently J1810+1714 \cite{Romani:2021xmb}, the gravitational waves detected from the NS binary merger GW170817 \cite{TheLIGOScientific:2017qsa,LIGOScientific:2018hze},  the NICER determination of the mass and radius of the PSR J0030+0451 \cite{Riley:2019yda,Miller:2019cac}, together with the determination of the radius of the PSR J0740+6620 from the joint analysis of data obtained by NICER and XMM-Newton \cite{Riley:2021pdl}. The total compatibility of the DDB set predictions with the  observations of NICER and of the LIGO/Virgo Collaboration indicates that more constraints are required to get more precise information on the high density EOS.  The presently existing constraints on the mass and radius are totally compatible with a composition restricted to nucleons and leptons.
{The lowest limit obtained for effective tidal defomability, $\tilde\Lambda_{q=1}\approx 382$ at 95\%CI, is above, but compatible, with the values $\tilde\Lambda\gtrsim 300$ and 242 determined in  \cite{Radice:2017lry} and \cite{Kiuchi_2019}, respectively, from the electromagnetic counterparts that followed up the GW170817 emission, i.e.  the gamma-ray burst
GRB170817A \cite{LIGOScientific:2017zic} and the electromagnetic
transient AT2017gfo \cite{LIGOScientific:2017ync}.}

{It has been shown that the generated set of models contain models with properties similar to TW \cite{Typel1999}, DD2 \cite{Typel2009} and DDME2 \cite{Lalazissis2005}, three DDH models frequently used in the literature, in particular, the last two. A common property of the DDB set of EOS  with these DDH models is the prediction that no nucleonic direct Urca occurs inside nucleonic NS, see \cite{Fortin2016,Fortin:2021umb}. This behavior requires the onset of hyperons inside the star to explain presently known cooling curves of the thermal evolution of nonmagnetized and nonrotating spherically-symmetric isolated  NS and accreting NS, \cite{Providencia:2018ywl,Fortin:2021umb}, which, however, may raise the problem of making the EOS too soft, not allowing for the existence of  2$M_\odot$ NSs. The effect of the onset of hyperons in models of the set DDB will be investigated in the future. 

\blu{It was verified that within the parametrization proposed for the couplings of the DDB set that maximum NS masses  obtained are  $\le2.51$~$M_\odot$, which is  just above the DD2 and DDME2 maximum mass and could  still be compatible with the low mass object of the binary merger that originated GW190814 \cite{LIGOScientific:2020zkf}. All  models with a mass above 2.48$M_\odot$ have an incompressibility $300\ge K_0\ge315$,  inside the range of values proposed in \cite{Stone:2014wza} where the incompressibility for infinite matter was determined from finite nuclei properties.}

\begin{acknowledgments}
This work was partially supported by national funds from FCT (Fundação para a Ciência e a Tecnologia, I.P, Portugal) under the Projects No. UID/\-FIS/\-04564/\-2019, No. UIDP/\-04564/\-2020, No. UIDB/\-04564/\-2020, and No. POCI-01-0145-FEDER-029912 with financial support from Science, Technology and Innovation, in its FEDER component, and by the FCT/MCTES budget through national funds (OE). BKA acknowledges partial  support from the
Department of Science and Technology, Government of India  with
grant no.  CRG/2021/000101. The authors acknowledge the Laboratory for Advanced Computing at University of Coimbra for providing {HPC} resources that have contributed to the research results reported within this paper, URL: \hyperlink{https://www.uc.pt/lca}{https://www.uc.pt/lca}.
\end{acknowledgments}

\blu{{{\it Data:} We are publicly releasing five tabulated  EOSs, namely DDBl, DDBm, DDBu1, DDBu2 and DDBx (see text of Section \ref{results} for details). We also release our entire sets of 14K NS matter EOS. All the EOSs are for NS core and starting baryon density is 0.04 fm$^{-3}$. One has to add their own choice of crust EOS to it for the calculation of star properties. The uncertainty in star properties for the choice of different crust has been discussed in Section \ref{model} of this manuscript.  All the EOS tables are available at \href{https://github.com/tuhinucpt/DDH_EOS}{\bf (https://github.com/tuhinucpt/DDH\_EOS)}.
The nuclear matter and NS properties for our five DDB EOSs can be found in the supplementary material.}}

\bibliography{main_R1}{}

\begin{thebibliography}{}
\expandafter\ifx\csname natexlab\endcsname\relax\def\natexlab#1{#1}\fi
\providecommand{\url}[1]{\href{#1}{#1}}
\providecommand{\dodoi}[1]{doi:~\href{http://doi.org/#1}{\nolinkurl{#1}}}
\providecommand{\doeprint}[1]{\href{http://ascl.net/#1}{\nolinkurl{http://ascl.net/#1}}}
\providecommand{\doarXiv}[1]{\href{https://arxiv.org/abs/#1}{\nolinkurl{https://arxiv.org/abs/#1}}}

\bibitem[{Abbott {et~al.}(2017{\natexlab{a}})}]{LIGOScientific:2017zic}
Abbott, B.~P., {et~al.} 2017{\natexlab{a}}, Astrophys. J. Lett., 848, L13,
  \dodoi{10.3847/2041-8213/aa920c}

\bibitem[{Abbott {et~al.}(2017{\natexlab{b}})}]{LIGOScientific:2017ync}
---. 2017{\natexlab{b}}, Astrophys. J. Lett., 848, L12,
  \dodoi{10.3847/2041-8213/aa91c9}

\bibitem[{Abbott {et~al.}(2017{\natexlab{c}})}]{TheLIGOScientific:2017qsa}
---. 2017{\natexlab{c}}, Phys. Rev. Lett., 119, 161101,
  \dodoi{10.1103/PhysRevLett.119.161101}

\bibitem[{Abbott {et~al.}(2018)}]{LIGOScientific:2018cki}
---. 2018, Phys. Rev. Lett., 121, 161101,
  \dodoi{10.1103/PhysRevLett.121.161101}

\bibitem[{Abbott {et~al.}(2019)}]{LIGOScientific:2018hze}
---. 2019, Phys. Rev. X, 9, 011001, \dodoi{10.1103/PhysRevX.9.011001}

\bibitem[{Abbott {et~al.}(2020)}]{LIGOScientific:2020zkf}
Abbott, R., {et~al.} 2020, Astrophys. J. Lett., 896, L44,
  \dodoi{10.3847/2041-8213/ab960f}

\bibitem[{Alford {et~al.}(2013)Alford, Han, \& Prakash}]{Alford:2013aca}
Alford, M.~G., Han, S., \& Prakash, M. 2013, Phys. Rev. D, 88, 083013,
  \dodoi{10.1103/PhysRevD.88.083013}

\bibitem[{Annala {et~al.}(2021)Annala, Gorda, Katerini, Kurkela, N\"attil\"a,
  Paschalidis, \& Vuorinen}]{Annala:2021gom}
Annala, E., Gorda, T., Katerini, E., {et~al.} 2021.
\newblock \doarXiv{2105.05132}

\bibitem[{Annala {et~al.}(2020)Annala, Gorda, Kurkela, N\"attil\"a, \&
  Vuorinen}]{Annala2019}
Annala, E., Gorda, T., Kurkela, A., N\"attil\"a, J., \& Vuorinen, A. 2020,
  Nature Phys., 16, 907, \dodoi{10.1038/s41567-020-0914-9}

\bibitem[{{Antoniadis} {et~al.}(2013){Antoniadis}, {Freire}, {Wex}, {Tauris},
  {Lynch}, {van Kerkwijk}, {Kramer}, {Bassa}, {Dhillon}, {Driebe}, {Hessels},
  {Kaspi}, {Kondratiev}, {Langer}, {Marsh}, {McLaughlin}, {Pennucci}, {Ransom},
  {Stairs}, {van Leeuwen}, {Verbiest}, \& {Whelan}}]{Antoniadis2013}
{Antoniadis}, J., {Freire}, P.~C.~C., {Wex}, N., {et~al.} 2013, Science, 340,
  448, \dodoi{10.1126/science.1233232}

\bibitem[{Arzoumanian {et~al.}(2018)}]{Arzoumanian2017}
Arzoumanian, Z., {et~al.} 2018, Astrophys. J. Suppl., 235, 37,
  \dodoi{10.3847/1538-4365/aab5b0}

\bibitem[{Ashton {et~al.}(2019)}]{Ashton2019}
Ashton, G., {et~al.} 2019, Astrophys. J. Suppl., 241, 27,
  \dodoi{10.3847/1538-4365/ab06fc}

\bibitem[{Avancini {et~al.}(2009)Avancini, Brito, Marinelli, Menezes,
  de~Moraes, Providencia, \& Santos}]{Avancini2008}
Avancini, S.~S., Brito, L., Marinelli, J.~R., {et~al.} 2009, Phys. Rev. C, 79,
  035804, \dodoi{10.1103/PhysRevC.79.035804}

\bibitem[{Avancini {et~al.}(2008)Avancini, Menezes, Alloy, Marinelli, Moraes,
  \& Providencia}]{Avancini:2008zz}
Avancini, S.~S., Menezes, D.~P., Alloy, M.~D., {et~al.} 2008, Phys. Rev. C, 78,
  015802, \dodoi{10.1103/PhysRevC.78.015802}

\bibitem[{{Baade} \& {Zwicky}(1934)}]{Baade1934a}
{Baade}, W., \& {Zwicky}, F. 1934, Proceedings of the National Academy of
  Science, 20, 259, \dodoi{10.1073/pnas.20.5.259}

\bibitem[{Baade \& Zwicky(1934)}]{Baade1934b}
Baade, W., \& Zwicky, F. 1934, Phys. Rev., 46, 76,
  \dodoi{10.1103/PhysRev.46.76.2}

\bibitem[{Bedaque \& Steiner(2015)}]{Bedaque:2014sqa}
Bedaque, P., \& Steiner, A.~W. 2015, Phys. Rev. Lett., 114, 031103,
  \dodoi{10.1103/PhysRevLett.114.031103}

\bibitem[{Boguta \& Bodmer(1977)}]{Boguta1977}
Boguta, J., \& Bodmer, A.~R. 1977, Nucl. Phys. A, 292, 413,
  \dodoi{10.1016/0375-9474(77)90626-1}

\bibitem[{{Brecher}(1999)}]{Brecher1999}
{Brecher}, K. 1999, in American Astronomical Society Meeting Abstracts, Vol.
  195, American Astronomical Society Meeting Abstracts, 130.05

\bibitem[{Brockmann \& Machleidt(1990)}]{Brockmann1990}
Brockmann, R., \& Machleidt, R. 1990, Phys. Rev. C, 42, 1965,
  \dodoi{10.1103/PhysRevC.42.1965}

\bibitem[{Buchner(2021)}]{buchner2021nested}
Buchner, J. 2021, Nested Sampling Methods.
\newblock \doarXiv{2101.09675}

\bibitem[{Buchner {et~al.}(2014)Buchner, Georgakakis, Nandra, Hsu, Rangel,
  Brightman, Merloni, Salvato, Donley, \& Kocevski}]{Buchner:2014nha}
Buchner, J., Georgakakis, A., Nandra, K., {et~al.} 2014, Astron. Astrophys.,
  564, A125, \dodoi{10.1051/0004-6361/201322971}

\bibitem[{Carriere {et~al.}(2003)Carriere, Horowitz, \&
  Piekarewicz}]{Carriere:2002bx}
Carriere, J., Horowitz, C.~J., \& Piekarewicz, J. 2003, Astrophys. J., 593,
  463, \dodoi{10.1086/376515}

\bibitem[{Constantinou {et~al.}(2015)Constantinou, Muccioli, Prakash, \&
  Lattimer}]{Constantinou2015}
Constantinou, C., Muccioli, B., Prakash, M., \& Lattimer, J.~M. 2015, Phys.
  Rev. C, 92, 025801, \dodoi{10.1103/PhysRevC.92.025801}

\bibitem[{Danielewicz {et~al.}(2002)Danielewicz, Lacey, \&
  Lynch}]{Danielewicz2002pu}
Danielewicz, P., Lacey, R., \& Lynch, W.~G. 2002, Science, 298, 1592,
  \dodoi{10.1126/science.1078070}

\bibitem[{Danielewicz \& Lee(2014)}]{Danielewicz:2013upa}
Danielewicz, P., \& Lee, J. 2014, Nucl. Phys. A, 922, 1,
  \dodoi{10.1016/j.nuclphysa.2013.11.005}

\bibitem[{de~Tovar {et~al.}(2021)de~Tovar, Ferreira, \&
  Provid\^encia}]{Tovar2021}
de~Tovar, P.~B., Ferreira, M., \& Provid\^encia, C. 2021, Phys. Rev. D, 104,
  123036, \dodoi{10.1103/PhysRevD.104.123036}

\bibitem[{Demorest {et~al.}(2010)Demorest, Pennucci, Ransom, Roberts, \&
  Hessels}]{Demorest2010}
Demorest, P., Pennucci, T., Ransom, S., Roberts, M., \& Hessels, J. 2010,
  Nature, 467, 1081, \dodoi{10.1038/nature09466}

\bibitem[{Drischler {et~al.}(2016)Drischler, Hebeler, \&
  Schwenk}]{Drischler2015}
Drischler, C., Hebeler, K., \& Schwenk, A. 2016, Phys. Rev. C, 93, 054314,
  \dodoi{10.1103/PhysRevC.93.054314}

\bibitem[{Drischler {et~al.}(2019)Drischler, Hebeler, \&
  Schwenk}]{Drischler:2017wtt}
---. 2019, Phys. Rev. Lett., 122, 042501,
  \dodoi{10.1103/PhysRevLett.122.042501}

\bibitem[{Dutra {et~al.}(2014)Dutra, Louren\c{c}o, Avancini, Carlson, Delfino,
  Menezes, Provid\^encia, Typel, \& Stone}]{Dutra:2014qga}
Dutra, M., Louren\c{c}o, O., Avancini, S.~S., {et~al.} 2014, Phys. Rev. C, 90,
  055203, \dodoi{10.1103/PhysRevC.90.055203}

\bibitem[{Essick {et~al.}(2020)Essick, Landry, \& Holz}]{Essick2019}
Essick, R., Landry, P., \& Holz, D.~E. 2020, Phys. Rev. D, 101, 063007,
  \dodoi{10.1103/PhysRevD.101.063007}

\bibitem[{Essick {et~al.}(2021{\natexlab{a}})Essick, Landry, Schwenk, \&
  Tews}]{Essick:2021ezp}
Essick, R., Landry, P., Schwenk, A., \& Tews, I. 2021{\natexlab{a}}, Phys. Rev.
  C, 104, 065804, \dodoi{10.1103/PhysRevC.104.065804}

\bibitem[{Essick {et~al.}(2021{\natexlab{b}})Essick, Tews, Landry, \&
  Schwenk}]{Essick2021}
Essick, R., Tews, I., Landry, P., \& Schwenk, A. 2021{\natexlab{b}}, Phys. Rev.
  Lett., 127, 192701, \dodoi{10.1103/PhysRevLett.127.192701}

\bibitem[{Fattoyev {et~al.}(2020)Fattoyev, Horowitz, Piekarewicz, \&
  Reed}]{Fattoyev:2020cws}
Fattoyev, F.~J., Horowitz, C.~J., Piekarewicz, J., \& Reed, B. 2020, Phys. Rev.
  C, 102, 065805, \dodoi{10.1103/PhysRevC.102.065805}

\bibitem[{Ferreira {et~al.}(2021)Ferreira, C\^amara~Pereira, \&
  Provid\^encia}]{Ferreira:2021osk}
Ferreira, M., C\^amara~Pereira, R., \& Provid\^encia, C. 2021, Phys. Rev. D,
  103, 123020, \dodoi{10.1103/PhysRevD.103.123020}

\bibitem[{Ferreira \& Provid\^encia(2021)}]{Ferreira:2021pni}
Ferreira, M., \& Provid\^encia, C. 2021, Phys. Rev. D, 104, 063006,
  \dodoi{10.1103/PhysRevD.104.063006}

\bibitem[{Ferreira \& Providência(2021)}]{Ferreira2021}
Ferreira, M., \& Providência, C. 2021, Journal of Cosmology and Astroparticle
  Physics, 2021, 011, \dodoi{10.1088/1475-7516/2021/07/011}

\bibitem[{Fonseca {et~al.}(2016)}]{Fonseca2016}
Fonseca, E., {et~al.} 2016, Astrophys. J., 832, 167,
  \dodoi{10.3847/0004-637X/832/2/167}

\bibitem[{Fonseca {et~al.}(2021)}]{Fonseca:2021wxt}
---. 2021, Astrophys. J. Lett., 915, L12, \dodoi{10.3847/2041-8213/ac03b8}

\bibitem[{Fortin {et~al.}(2016)Fortin, Providencia, Raduta, Gulminelli, Zdunik,
  Haensel, \& Bejger}]{Fortin2016}
Fortin, M., Providencia, C., Raduta, A.~R., {et~al.} 2016, Phys. Rev. C, 94,
  035804, \dodoi{10.1103/PhysRevC.94.035804}

\bibitem[{Fortin {et~al.}(2020)Fortin, Raduta, Avancini, \&
  Provid\^encia}]{Fortin:2020qin}
Fortin, M., Raduta, A.~R., Avancini, S., \& Provid\^encia, C. 2020, Phys. Rev.
  D, 101, 034017, \dodoi{10.1103/PhysRevD.101.034017}

\bibitem[{Fortin {et~al.}(2021)Fortin, Raduta, Avancini, \&
  Provid\^encia}]{Fortin:2021umb}
---. 2021, Phys. Rev. D, 103, 083004, \dodoi{10.1103/PhysRevD.103.083004}

\bibitem[{Fritz \& Muther(1994)}]{Fritz:1994ww}
Fritz, R., \& Muther, H. 1994, Phys. Rev. C, 49, 633,
  \dodoi{10.1103/PhysRevC.49.633}

\bibitem[{Fuchs {et~al.}(1995)Fuchs, Lenske, \& Wolter}]{Fuchs:1995as}
Fuchs, C., Lenske, H., \& Wolter, H.~H. 1995, Phys. Rev. C, 52, 3043,
  \dodoi{10.1103/PhysRevC.52.3043}

\bibitem[{Furnstahl {et~al.}(2015)Furnstahl, Klco, Phillips, \&
  Wesolowski}]{Furnstahl:2015rha}
Furnstahl, R.~J., Klco, N., Phillips, D.~R., \& Wesolowski, S. 2015, Phys. Rev.
  C, 92, 024005, \dodoi{10.1103/PhysRevC.92.024005}

\bibitem[{{Glendenning}(1996)}]{book.Glendenning1996}
{Glendenning}, N.~K. 1996, {Compact Stars}

\bibitem[{Haddad \& Weigel(1993)}]{Haddad:1993zz}
Haddad, S., \& Weigel, M.~K. 1993, Phys. Rev. C, 48, 2740,
  \dodoi{10.1103/PhysRevC.48.2740}

\bibitem[{{Haensel} {et~al.}(2007){Haensel}, {Potekhin}, \&
  {Yakovlev}}]{book.Haensel2007}
{Haensel}, P., {Potekhin}, A.~Y., \& {Yakovlev}, D.~G. 2007, {Neutron Stars 1 :
  Equation of State and Structure}, Vol. 326

\bibitem[{Han {et~al.}(2021)Han, Jiang, Tang, \& Fan}]{Han:2021kjx}
Han, M.-Z., Jiang, J.-L., Tang, S.-P., \& Fan, Y.-Z. 2021, Astrophys. J., 919,
  11, \dodoi{10.3847/1538-4357/ac11f8}

\bibitem[{Hebeler {et~al.}(2013)Hebeler, Lattimer, Pethick, \&
  Schwenk}]{Hebeler2013}
Hebeler, K., Lattimer, J.~M., Pethick, C.~J., \& Schwenk, A. 2013, Astrophys.
  J., 773, 11, \dodoi{10.1088/0004-637X/773/1/11}

\bibitem[{Hebeler \& Schwenk(2010)}]{Hebeler:2009iv}
Hebeler, K., \& Schwenk, A. 2010, Phys. Rev. C, 82, 014314,
  \dodoi{10.1103/PhysRevC.82.014314}

\bibitem[{{Hewish} {et~al.}(1968){Hewish}, {Bell}, {Pilkington}, {Scott}, \&
  {Collins}}]{Bell1968}
{Hewish}, A., {Bell}, S.~J., {Pilkington}, J.~D.~H., {Scott}, P.~F., \&
  {Collins}, R.~A. 1968, \nat, 217, 709, \dodoi{10.1038/217709a0}

\bibitem[{Hinderer(2008)}]{Hinderer2008}
Hinderer, T. 2008, Astrophys. J., 677, 1216, \dodoi{10.1086/533487}

\bibitem[{Huang {et~al.}(2020)Huang, Hu, Zhang, \& Shen}]{Huang:2020cab}
Huang, K., Hu, J., Zhang, Y., \& Shen, H. 2020, Astrophys. J., 904, 39,
  \dodoi{10.3847/1538-4357/abbb37}

\bibitem[{Imam {et~al.}(2021)Imam, Patra, Mondal, Malik, \& Agrawal}]{Imam2021}
Imam, S. M.~A., Patra, N.~K., Mondal, C., Malik, T., \& Agrawal, B.~K. 2021.
\newblock \doarXiv{2110.15776}

\bibitem[{Kiuchi {et~al.}(2019)Kiuchi, Kyutoku, Shibata, \&
  Taniguchi}]{Kiuchi_2019}
Kiuchi, K., Kyutoku, K., Shibata, M., \& Taniguchi, K. 2019, The Astrophysical
  Journal, 876, L31, \dodoi{10.3847/2041-8213/ab1e45}

\bibitem[{Kurkela {et~al.}(2014)Kurkela, Fraga, Schaffner-Bielich, \&
  Vuorinen}]{Kurkela:2014vha}
Kurkela, A., Fraga, E.~S., Schaffner-Bielich, J., \& Vuorinen, A. 2014,
  Astrophys. J., 789, 127, \dodoi{10.1088/0004-637X/789/2/127}

\bibitem[{Kurkela {et~al.}(2010)Kurkela, Romatschke, \& Vuorinen}]{Kurkela2009}
Kurkela, A., Romatschke, P., \& Vuorinen, A. 2010, Phys. Rev. D, 81, 105021,
  \dodoi{10.1103/PhysRevD.81.105021}

\bibitem[{Lalazissis {et~al.}(2005)Lalazissis, Niksic, Vretenar, \&
  Ring}]{Lalazissis2005}
Lalazissis, G.~A., Niksic, T., Vretenar, D., \& Ring, P. 2005, Phys. Rev. C,
  71, 024312, \dodoi{10.1103/PhysRevC.71.024312}

\bibitem[{Landry {et~al.}(2020)Landry, Essick, \&
  Chatziioannou}]{Landry:2020vaw}
Landry, P., Essick, R., \& Chatziioannou, K. 2020, Phys. Rev. D, 101, 123007,
  \dodoi{10.1103/PhysRevD.101.123007}

\bibitem[{Lattimer \& Lim(2013)}]{Lattimer2013}
Lattimer, J.~M., \& Lim, Y. 2013, Astrophys. J., 771, 51,
  \dodoi{10.1088/0004-637X/771/1/51}

\bibitem[{Lattimer {et~al.}(1991)Lattimer, Prakash, Pethick, \&
  Haensel}]{Lattimer:1991ib}
Lattimer, J.~M., Prakash, M., Pethick, C.~J., \& Haensel, P. 1991, Phys. Rev.
  Lett., 66, 2701, \dodoi{10.1103/PhysRevLett.66.2701}

\bibitem[{Lenske \& Fuchs(1995)}]{Lenske:1995wyj}
Lenske, H., \& Fuchs, C. 1995, Phys. Lett. B, 345, 355,
  \dodoi{10.1016/0370-2693(94)01664-X}

\bibitem[{Li {et~al.}(2019)Li, Krastev, Wen, \& Zhang}]{Li:2019xxz}
Li, B.-A., Krastev, P.~G., Wen, D.-H., \& Zhang, N.-B. 2019, Eur. Phys. J. A,
  55, 117, \dodoi{10.1140/epja/i2019-12780-8}

\bibitem[{Lindblom \& Indik(2012)}]{Lindblom2012}
Lindblom, L., \& Indik, N.~M. 2012, Phys. Rev. D, 86, 084003,
  \dodoi{10.1103/PhysRevD.86.084003}

\bibitem[{Loh(1996)}]{LHS1}
Loh, W.-L. 1996, The Annals of Statistics, 24, 2058 ,
  \dodoi{10.1214/aos/1069362310}

\bibitem[{Lope~Oter {et~al.}(2019)Lope~Oter, Windisch, Llanes-Estrada, \&
  Alford}]{LopeOter:2019pcq}
Lope~Oter, E., Windisch, A., Llanes-Estrada, F.~J., \& Alford, M. 2019, J.
  Phys. G, 46, 084001, \dodoi{10.1088/1361-6471/ab2567}

\bibitem[{Lopes(2021)}]{Lopes:2020xlf}
Lopes, L.~L. 2021, EPL, 134, 52001, \dodoi{10.1209/0295-5075/134/52001}

\bibitem[{Lynn {et~al.}(2016)Lynn, Tews, Carlson, Gandolfi, Gezerlis, Schmidt,
  \& Schwenk}]{Lynn:2015jua}
Lynn, J.~E., Tews, I., Carlson, J., {et~al.} 2016, Phys. Rev. Lett., 116,
  062501, \dodoi{10.1103/PhysRevLett.116.062501}

\bibitem[{Marcos {et~al.}(1989)Marcos, Niembro, Lopez-Quelle, Van~Giai, \&
  Malfliet}]{Marcos:1989zz}
Marcos, S., Niembro, R., Lopez-Quelle, M., Van~Giai, N., \& Malfliet, R. 1989,
  Phys. Rev. C, 39, 1134, \dodoi{10.1103/PhysRevC.39.1134}

\bibitem[{Margueron {et~al.}(2018{\natexlab{a}})Margueron, Hoffmann~Casali, \&
  Gulminelli}]{Margueron2018a}
Margueron, J., Hoffmann~Casali, R., \& Gulminelli, F. 2018{\natexlab{a}}, Phys.
  Rev., C97, 025805, \dodoi{10.1103/PhysRevC.97.025805}

\bibitem[{Margueron {et~al.}(2018{\natexlab{b}})Margueron, Hoffmann~Casali, \&
  Gulminelli}]{Margueron2018b}
---. 2018{\natexlab{b}}, Phys. Rev., C97, 025806,
  \dodoi{10.1103/PhysRevC.97.025806}

\bibitem[{Miller {et~al.}(2019)}]{Miller:2019cac}
Miller, M.~C., {et~al.} 2019, Astrophys. J. Lett., 887, L24,
  \dodoi{10.3847/2041-8213/ab50c5}

\bibitem[{Miller {et~al.}(2021)}]{Miller:2021qha}
---. 2021, Astrophys. J. Lett., 918, L28, \dodoi{10.3847/2041-8213/ac089b}

\bibitem[{Mondal \& Gulminelli(2021)}]{Mondal2021}
Mondal, C., \& Gulminelli, F. 2021.
\newblock \doarXiv{2111.04520}

\bibitem[{Most {et~al.}(2018)Most, Weih, Rezzolla, \&
  Schaffner-Bielich}]{Most:2018hfd}
Most, E.~R., Weih, L.~R., Rezzolla, L., \& Schaffner-Bielich, J. 2018, Phys.
  Rev. Lett., 120, 261103, \dodoi{10.1103/PhysRevLett.120.261103}

\bibitem[{Moustakidis {et~al.}(2017)Moustakidis, Gaitanos, Margaritis, \&
  Lalazissis}]{Moustakidis:2016sab}
Moustakidis, C.~C., Gaitanos, T., Margaritis, C., \& Lalazissis, G.~A. 2017,
  Phys. Rev. C, 95, 045801, \dodoi{10.1103/PhysRevC.95.045801}

\bibitem[{Mueller \& Serot(1996)}]{Mueller1996}
Mueller, H., \& Serot, B.~D. 1996, Nucl. Phys. A, 606, 508,
  \dodoi{10.1016/0375-9474(96)00187-X}

\bibitem[{Oppenheimer \& Volkoff(1939)}]{TOV2}
Oppenheimer, J.~R., \& Volkoff, G.~M. 1939, Phys. Rev., 55, 374,
  \dodoi{10.1103/PhysRev.55.374}

\bibitem[{Pais \& Provid\^encia(2016)}]{Pais:2016xiu}
Pais, H., \& Provid\^encia, C. 2016, Phys. Rev. C, 94, 015808,
  \dodoi{10.1103/PhysRevC.94.015808}

\bibitem[{Provid\^encia {et~al.}(2018)Provid\^encia, Fortin, Pais, \&
  Rabhi}]{Providencia:2018ywl}
Provid\^encia, C., Fortin, M., Pais, H., \& Rabhi, A. 2018,
  \dodoi{10.3389/fspas.2019.00013}

\bibitem[{Radice {et~al.}(2018)Radice, Perego, Zappa, \&
  Bernuzzi}]{Radice:2017lry}
Radice, D., Perego, A., Zappa, F., \& Bernuzzi, S. 2018, Astrophys. J. Lett.,
  852, L29, \dodoi{10.3847/2041-8213/aaa402}

\bibitem[{Rather {et~al.}(2021)Rather, Usmani, \& Patra}]{Rather:2020gja}
Rather, I.~A., Usmani, A.~A., \& Patra, S.~K. 2021, Nucl. Phys. A, 1010,
  122189, \dodoi{10.1016/j.nuclphysa.2021.122189}

\bibitem[{Reed \& Horowitz(2020)}]{Reed:2019ezm}
Reed, B., \& Horowitz, C.~J. 2020, Phys. Rev. C, 101, 045803,
  \dodoi{10.1103/PhysRevC.101.045803}

\bibitem[{Reed {et~al.}(2021)Reed, Fattoyev, Horowitz, \& Piekarewicz}]{PREX2}
Reed, B.~T., Fattoyev, F.~J., Horowitz, C.~J., \& Piekarewicz, J. 2021, Phys.
  Rev. Lett., 126, 172503, \dodoi{10.1103/PhysRevLett.126.172503}

\bibitem[{Rezzolla {et~al.}(2018)Rezzolla, Pizzochero, Jones, Rea, \&
  Vida\~na}]{Rezzolla:2018jee}
Rezzolla, L., Pizzochero, P., Jones, D.~I., Rea, N., \& Vida\~na, I., eds.
  2018, {The Physics and Astrophysics of Neutron Stars}, Vol. 457 (Springer),
  \dodoi{10.1007/978-3-319-97616-7}

\bibitem[{Riley {et~al.}(2019)}]{Riley:2019yda}
Riley, T.~E., {et~al.} 2019, Astrophys. J. Lett., 887, L21,
  \dodoi{10.3847/2041-8213/ab481c}

\bibitem[{Riley {et~al.}(2021)}]{Riley:2021pdl}
---. 2021, Astrophys. J. Lett., 918, L27, \dodoi{10.3847/2041-8213/ac0a81}

\bibitem[{Romani {et~al.}(2021)Romani, Kandel, Filippenko, Brink, \&
  Zheng}]{Romani:2021xmb}
Romani, R.~W., Kandel, D., Filippenko, A.~V., Brink, T.~G., \& Zheng, W. 2021,
  Astrophys. J. Lett., 908, L46, \dodoi{10.3847/2041-8213/abe2b4}

\bibitem[{Serot \& Walecka(1986)}]{Serot1984}
Serot, B.~D., \& Walecka, J.~D. 1986, Adv. Nucl. Phys., 16, 1

\bibitem[{{Shlomo, S.} {et~al.}(2006){Shlomo, S.}, {Kolomietz, V. M.}, \&
  {Col\`o, G.}}]{Shlomo2006}
{Shlomo, S.}, {Kolomietz, V. M.}, \& {Col\`o, G.} 2006, Eur. Phys. J. A, 30,
  23, \dodoi{10.1140/epja/i2006-10100-3}

\bibitem[{{Skilling}(2004)}]{Skilling2004}
{Skilling}, J. 2004, in American Institute of Physics Conference Series, Vol.
  735, Bayesian Inference and Maximum Entropy Methods in Science and
  Engineering: 24th International Workshop on Bayesian Inference and Maximum
  Entropy Methods in Science and Engineering, ed. R.~{Fischer}, R.~{Preuss}, \&
  U.~V. {Toussaint}, 395--405, \dodoi{10.1063/1.1835238}

\bibitem[{Speagle(2020)}]{Speagle2019}
Speagle, J.~S. 2020, Mon. Not. Roy. Astron. Soc., 493, 3132,
  \dodoi{10.1093/mnras/staa278}

\bibitem[{Steiner {et~al.}(2005)Steiner, Prakash, Lattimer, \&
  Ellis}]{Steiner2004}
Steiner, A.~W., Prakash, M., Lattimer, J.~M., \& Ellis, P.~J. 2005, Phys.
  Rept., 411, 325, \dodoi{10.1016/j.physrep.2005.02.004}

\bibitem[{Stone {et~al.}(2014)Stone, Stone, \& Moszkowski}]{Stone:2014wza}
Stone, J.~R., Stone, N.~J., \& Moszkowski, S.~A. 2014, Phys. Rev. C, 89,
  044316, \dodoi{10.1103/PhysRevC.89.044316}

\bibitem[{Taninah {et~al.}(2020)Taninah, Agbemava, Afanasjev, \&
  Ring}]{Taninah:2019cku}
Taninah, A., Agbemava, S.~E., Afanasjev, A.~V., \& Ring, P. 2020, Phys. Lett.
  B, 800, 135065, \dodoi{10.1016/j.physletb.2019.135065}

\bibitem[{Ter~Haar \& Malfliet(1987)}]{TerHaar1987}
Ter~Haar, B., \& Malfliet, R. 1987, Phys. Rept., 149, 207,
  \dodoi{10.1016/0370-1573(87)90085-8}

\bibitem[{Tews {et~al.}(2018)Tews, Carlson, Gandolfi, \& Reddy}]{Tews:2018kmu}
Tews, I., Carlson, J., Gandolfi, S., \& Reddy, S. 2018, Astrophys. J., 860,
  149, \dodoi{10.3847/1538-4357/aac267}

\bibitem[{Tews {et~al.}(2013)Tews, Kr\"uger, Hebeler, \& Schwenk}]{Tews:2012fj}
Tews, I., Kr\"uger, T., Hebeler, K., \& Schwenk, A. 2013, Phys. Rev. Lett.,
  110, 032504, \dodoi{10.1103/PhysRevLett.110.032504}

\bibitem[{Tews {et~al.}(2017)Tews, Lattimer, Ohnishi, \&
  Kolomeitsev}]{Tews:2016jhi}
Tews, I., Lattimer, J.~M., Ohnishi, A., \& Kolomeitsev, E.~E. 2017, Astrophys.
  J., 848, 105, \dodoi{10.3847/1538-4357/aa8db9}

\bibitem[{Thi {et~al.}(2021)Thi, Mondal, \& Gulminelli}]{Thi:2021jhz}
Thi, H.~D., Mondal, C., \& Gulminelli, F. 2021, Universe, 7, 373,
  \dodoi{10.3390/universe7100373}

\bibitem[{Todd-Rutel \& Piekarewicz(2005)}]{Todd-Rutel2005}
Todd-Rutel, B.~G., \& Piekarewicz, J. 2005, Phys. Rev. Lett., 95, 122501,
  \dodoi{10.1103/PhysRevLett.95.122501}

\bibitem[{Tolman(1939)}]{TOV1}
Tolman, R.~C. 1939, Phys. Rev., 55, 364, \dodoi{10.1103/PhysRev.55.364}

\bibitem[{Typel {et~al.}(2010)Typel, Ropke, Klahn, Blaschke, \&
  Wolter}]{Typel2009}
Typel, S., Ropke, G., Klahn, T., Blaschke, D., \& Wolter, H.~H. 2010, Phys.
  Rev. C, 81, 015803, \dodoi{10.1103/PhysRevC.81.015803}

\bibitem[{Typel \& Wolter(1999)}]{Typel1999}
Typel, S., \& Wolter, H.~H. 1999, Nucl. Phys. A, 656, 331,
  \dodoi{10.1016/S0375-9474(99)00310-3}

\bibitem[{Vidana {et~al.}(2009)Vidana, Providencia, Polls, \&
  Rios}]{Vidana2009}
Vidana, I., Providencia, C., Polls, A., \& Rios, A. 2009, Phys. Rev., C80,
  045806, \dodoi{10.1103/PhysRevC.80.045806}

\bibitem[{Wei {et~al.}(2020)Wei, Zhao, Wang, Geng, Sun, Niu, \&
  Long}]{Wei:2020kfb}
Wei, B., Zhao, Q., Wang, Z.-H., {et~al.} 2020, Chin. Phys. C, 44, 074107,
  \dodoi{10.1088/1674-1137/44/7/074107}

\bibitem[{Wesolowski {et~al.}(2016)Wesolowski, Klco, Furnstahl, Phillips, \&
  Thapaliya}]{Wesolowski:2015fqa}
Wesolowski, S., Klco, N., Furnstahl, R.~J., Phillips, D.~R., \& Thapaliya, A.
  2016, J. Phys. G, 43, 074001, \dodoi{10.1088/0954-3899/43/7/074001}

\bibitem[{Yakovlev {et~al.}(2013)Yakovlev, Haensel, Baym, \&
  Pethick}]{Yakovlev:2012rd}
Yakovlev, D.~G., Haensel, P., Baym, G., \& Pethick, C.~J. 2013, Phys. Usp., 56,
  289, \dodoi{10.3367/UFNe.0183.201303f.0307}

\bibitem[{Yakovlev {et~al.}(2001)Yakovlev, Kaminker, Gnedin, \&
  Haensel}]{Yakovlev:2000jp}
Yakovlev, D.~G., Kaminker, A.~D., Gnedin, O.~Y., \& Haensel, P. 2001, Phys.
  Rept., 354, 1, \dodoi{10.1016/S0370-1573(00)00131-9}

\bibitem[{Zhang \& Li(2019)}]{Zhang:2019fog}
Zhang, N.-B., \& Li, B.-A. 2019, Astrophys. J., 879, 99,
  \dodoi{10.3847/1538-4357/ab24cb}

\bibitem[{Zhang {et~al.}(2018)Zhang, Li, \& Xu}]{Zhang2018}
Zhang, N.-B., Li, B.-A., \& Xu, J. 2018, Astrophys. J., 859, 90,
  \dodoi{10.3847/1538-4357/aac027}

\end{thebibliography}
\bibliographystyle{aasjournal}


\end{document}